\documentclass[lettersize,journal]{IEEEtran}
\usepackage{amsmath,amsfonts}
\usepackage{algorithmic}
\usepackage{algorithm}
\usepackage{array}
\usepackage[caption=false,font=normalsize,labelfont=sf,textfont=sf]{subfig}
\usepackage{textcomp}
\usepackage{stfloats}
\usepackage{url}
\usepackage{hyperref}
\usepackage{verbatim}
\usepackage{graphicx}
\usepackage{cite}
\usepackage{makecell}
\usepackage{multirow}
\usepackage{color}
\hypersetup{hidelinks}

\begin{document}

\title{Identifying social bots via heterogeneous motifs based on Na\"ive Bayes model}

\author{Yijun~Ran, Jingjing~Xiao and Xiao-Ke~Xu
\thanks{This work was supported in part by the National Natural Science Foundation of China under Grant 62403062, Grant 62573061, and Grant 62173065, in part by the Beijing Natural Science Foundation under Grant 4242040, in part by the Fundamental Research Funds for the Central Universities under Grant 124330008, and Grant 123330009, in part by the Postdoctoral Fellowship Program of CPSF under Grant GZC20230281, in part by the Foundation of State Key Laboratory of Public Big Data under Grant PBD2023-19, and in part by the Beijing Social Science Foundation under Grant 21DTR040. (Corresponding author: Xiao-Ke Xu.)}

\IEEEcompsocitemizethanks{\IEEEcompsocthanksitem Yijun Ran is with the State Key Laboratory of Public Big Data, Guizhou University, Guiyang 550025, People's Republic of China, and School of Big Data and Computer Science, Guizhou Normal University, Guiyang 550025, People's Republic of China, and Center for Computational Communication Research, Beijing Normal University, Zhuhai 519087, People's Republic of China (e-mail: ranyij288@gmail.com).}

\IEEEcompsocitemizethanks{\IEEEcompsocthanksitem Jingjing Xiao is with the College of Information and Communication Engineering, Dalian Minzu University, Dalian 116600, People's Republic of China (e-mail: 18042662689@163.com).}

\IEEEcompsocitemizethanks{\IEEEcompsocthanksitem Xiao-Ke Xu is with the Center for Computational Communication Research, Beijing Normal University, Zhuhai 519087, People's Republic of China, and School of Journalism and Communication, Beijing Normal University, Beijing 100875, People's Republic of China (e-mail: xuxiaoke@foxmail.com).}}

\maketitle

\begin{abstract}
Identifying social bots has become a critical challenge due to their significant influence on social media ecosystems. Despite advancements in detection methods, most topology-based approaches insufficiently account for the heterogeneity of neighborhood preferences and lack a systematic theoretical foundation, relying instead on intuition and experience. Here, we propose a theoretical framework for detecting social bots utilizing heterogeneous motifs based on the Na\"ive Bayes model. Specifically, we refine homogeneous motifs into heterogeneous ones by incorporating node-label information, effectively capturing the heterogeneity of neighborhood preferences. Additionally, we systematically evaluate the contribution of different node pairs within heterogeneous motifs to the likelihood of a node being identified as a social bot. Furthermore, we mathematically quantify the maximum capability of each heterogeneous motif, enabling the estimation of its potential benefits. Comprehensive evaluations on four large, publicly available benchmarks confirm that our method surpasses state-of-the-art techniques, achieving superior performance across five evaluation metrics. Moreover, our results reveal that selecting motifs with the highest capability achieves detection performance comparable to using all heterogeneous motifs. Overall, our framework offers an effective and theoretically grounded solution for social bot detection, significantly enhancing cybersecurity measures in social networks.
\end{abstract}


\section{Introduction}
\IEEEPARstart{T}~he proliferation of social media platforms such as Facebook, Twitter (now X), TikTok, and Weibo has transformed how individuals communicate, share information, and form communities \cite{allen2024quantifying,notarmuzi2022universality}. However, this transformation has also given rise to the pervasive issue of social bots---automated accounts rather than real individuals that mimic human behavior on social network platforms \cite{chen2021neutral,shao2018spread}. Indeed, while some social bots serve benign purposes, many are involved in malicious activities \cite{cresci2020decade,latah2020detection}. These malicious bots can manipulate online public opinion \cite{bessi2016social}, distort information spreading \cite{stella2018bots}, and pose significant threats to both their neighbors and social network platforms \cite{ferrara2016rise}. The illusory truth effect, where repeated exposure to false information makes it seem more credible, is often exploited by social bots, amplifying their negative impact \cite{gonzalez2021bots}. Therefore, detecting and addressing social bots is essential for preserving the integrity of social networks and ensuring account security against potential cybersecurity threats.

The problem of social bot detection has garnered significant attention from researchers across multiple fields, including network science \cite{peng2024unsupervised,des2022detecting}, artificial intelligence \cite{feng2021botrgcn,feng2021satar,ilias2024multimodal}, and cybersecurity \cite{yang2013empirical,fazil2021deepsbd,arin2023deep}. Social bot detection methods are typically categorized into metadata-based, content-based, and topology-based \cite{feng2022twibot,huang2024cgnn,qiao2024dispelling}. Because account metadata \cite{yang2020scalable,yang2022botometer,moghaddam2022friendship}, including the number of followers, the number of favorites, and other profile characteristics, is usually directly available for a given social account, early significant efforts have been devoted to utilizing metadata features to identify social bots. With the development of computational tools, especially word embedding techniques powered by deep learning, many works have leveraged contents to design social bot detectors \cite{orabi2020detection,dehghan2023detecting}. Moreover, recent research has also focused on leveraging follow relationships to improve the accuracy and robustness of bot detection using Graph Neural Networks (GNNs), as GNNs can directly utilize network topology without prior knowledge of every component within the network \cite{feng2021botrgcn,feng2022heterogeneity,deng2022markov,shi2023rf,huang2024cgnn,qiao2024dispelling}.

For network topology, mainstream methods usually use computational tools from network science and deep learning for social bot detection. For example, UnDBot utilizes heterogeneous structural entropy to identify social bots by three social relationship metrics such as posting type distribution, posting influence, and follow-to-follower ratio \cite{peng2024unsupervised}. To overcome community challenges, BotRGCN constructs a heterogeneous graph from various relationships and applies relational graph convolutional neural networks for social bot detection \cite{feng2021botrgcn}. Additionally, to explore the relationship and influence heterogeneity, RGT leverages the relational graph transformer to improve bot detection performance \cite{feng2022heterogeneity}. Despite significant progress, the heterogeneity of neighborhood preferences is still inadequately explored. Furthermore, existing topology-based social bot detection methods rely heavily on intuition and experience, lacking a systematic theoretical framework. This gap highlights the need to develop a theoretical tool capable of quantifying the contribution of different node pairs within heterogeneous motifs for detecting social bots.

Here, we propose an interpretable and effective framework for social bot detection, utilizing heterogeneous network motifs based on the Na\"ive Bayes model. Network motifs are small, recurring subgraph patterns that occur significantly more frequently in a real-world network than in randomized networks \cite{milo2002network,alon2007network}. These motifs serve as fundamental building blocks of complex networks, offering valuable insights into the network structure and functional properties. Specifically, we begin by introducing homogeneous motifs, which are then refined into heterogeneous motifs by incorporating node-label information. Next, we use the Na\"ive Bayes model to quantify the contribution of different node pairs within heterogeneous motifs to the probability of a node being identified as a social bot. Lastly, we calculate the maximum capability of each heterogeneous motif for detecting social bots and demonstrate how this value can be used to evaluate detection performance when combining multiple heterogeneous motifs. The maximum capability represents the upper limit of a feature's effectiveness in identifying social bots. The main contributions of this work can be summarized as follows.

\begin{enumerate}
	\item To capture the heterogeneity of neighborhood preferences, we introduce heterogeneous motifs by enhancing homogeneous motifs with node-label information. This approach leverages structural information in social networks more effectively, highlighting the importance of understanding social relationships and neighborhood preferences in the context of a target node.
	
	\item To distinctly and theoretically quantify the contribution of different node pairs within heterogeneous motifs, we propose a novel framework based on the Na\"ive Bayes model for detecting social bots. A key advantage of this framework is its nonparametric nature, which eliminates the need for parameter tuning.
	
	\item To further explore the role of a heterogeneous motif, we quantify its maximum capability in social bot detection, offering valuable insights for feature selection. Importantly, the derived mathematical expression for the maximum capability of a feature is widely applicable across various machine learning algorithms, providing a theoretical foundation for interpretability.
	
	\item Through rigorous evaluation on four large-scale, publicly available social bot detection benchmarks, our method demonstrates clear and consistent superiority over existing state-of-the-art approaches. It achieves the best performance under five standard evaluation metrics, highlighting both its robustness and generalizability.
	
\end{enumerate}

The rest of this study is organized as follows. In Section \ref{section2}, we review the related work of social bot detection in detail. Section \ref{section4} gives the performance metrics and feature selection techniques. Section \ref{section3} presents the concepts of homogeneous and heterogeneous motifs and outlines a theoretical framework grounded in the Na\"ive Bayes model. In Section \ref{section5}, we conduct experiments and report the main results. Finally, the conclusion and discussion of this study are presented in Section \ref{section6}.

\section{Related work}
\label{section2}
The goal of social bot detection is to differentiate between human accounts and bots by primarily leveraging account metadata, textual content, and network topology. This information is typically analyzed through heuristic approaches, natural language processing, and GNNs. Here, we provide a detailed overview of the most relevant studies employing these methods.

\subsection{Social Bot Detection Methods}
\label{section21}
In metadata-based methods, account metadata is typically extracted directly from the account object retrieved via the API. This metadata often provides both numerical and binary features for analysis. During the initial phase of social bot detection, researchers primarily use heuristic approaches based on clear and recognizable metadata patterns. These patterns include an unusually high following-to-follower ratio and a short account lifespan, which are frequently observed in social bots \cite{yang2022botometer}. With the availability of various features extracted from metadata, traditional classification algorithms like Support Vector Machines \cite{abreu2020twitter}, Logistic Regression \cite{stella2018bots}, Random Forests \cite{yang2020scalable,yang2022botometer}, and K-means \cite{miller2014twitter,wu2022twitter} have been employed to classify accounts as bots or humans. Recently, most bot detection methods based on deep learning frameworks have also leveraged metadata to enhance the representation of account profiles \cite{feng2021botrgcn,feng2021satar,hayawi2022deeprobot}.

In content-based methods, textual content encompasses a wealth of information, including the text itself and critical elements like time series data, hashtags, symbols, and URLs, which provide a comprehensive understanding of account behavior and interactions. Similarly, early efforts to detect social bots using textual content largely focus on capturing and analyzing temporal activity patterns to identify coordinated or anomalous behaviors among accounts \cite{cresci2017social,al2017leveraging}. For instance, the study \cite{cresci2017social} introduces a method inspired by biological DNA, encoding account behaviors such as posting, retweeting, and commenting into a sequence of characters, then uses the longest common subsequence analysis to detect social bots. With the advancement of deep learning, numerous natural language processing techniques including word embeddings \cite{feng2021botrgcn,feng2021satar} and pre-trained language models \cite{raffel2020exploring}, have been adopted to learn feature embeddings from textual content for bot detection. 

Recent advancements and widespread adoption of large language models (LLMs) have enabled bots to generate human-like account metadata and textual content seamlessly \cite{yang2023anatomy}. Consequently, detection methods that rely on metadata and content may face significant challenges in identifying these bots.

\subsection{Topology-based Social Bot Detection Methods}
\label{section23}
Numerous user interaction patterns on social media, such as following (or follower), replying, and reposting relationships, can be effectively represented and analyzed through network topology. This captures the complexity and interconnectedness of user behaviors, providing valuable insights for social bot detection \cite{dehghan2023detecting,peng2024unsupervised}. Early works on bot (or Sybil) detection often relied on the assumption that influence, trust, or reputation could be spread as much as possible from one bot (or human account) to another bot (or human account) through social relationships. These works include SybilLimit \cite{yu2008sybillimit}, SybilDefender \cite{wei2012sybildefender}, SybilRank \cite{cao2012aiding}, SybilBelief \cite{gong2014sybilbelief}, SmartWalk \cite{liu2016smartwalk}, SybilWalk \cite{jia2017random} and SybilSCAR \cite{wang2018structure}.

Recently, GNNs have been applied to social bot detection, proving more effective in utilizing network topology than traditional machine learning algorithms \cite{feng2021botrgcn,feng2022heterogeneity,deng2022markov,shi2023rf,huang2024cgnn,qiao2024dispelling}. For example, Feng \textit{et al.} propose BotRGCN, which overcomes community challenges by constructing a heterogeneous network and utilizing relational graph convolutional networks \cite{feng2021botrgcn}. Similarly, Deng \textit{et al.} introduce MDGCN, which integrates conditional random fields and an adaptive reward Markov random field layer with graph convolutional networks to enhance expressive power \cite{deng2022markov}. To better adapt to diverse user communities, Liu \textit{et al.} propose BotMoE, a multi-modal Twitter bot detection framework that captures a community-aware mixture-of-experts mechanism to improve generalization and detection performance \cite{liu2023botmoe}. Subsequently, Shi \textit{et al.} design a graph neural network enhanced by random forest for social bot detection, effectively combining the strengths of both ensemble learning and GNNs \cite{shi2023rf}. Furthermore, they propose OS-GNN, an over-sampling framework that generates synthetic minority-class samples in the feature space without modifying the network structure \cite{shi2024over}. To capture higher-order neighborhood interactions, they also introduce NDE-GNN, a hypergraph-based model that fuses hypergraph and graph features via self-attention to capture neighborhood differences and improve detection performance \cite{shi2024neighborhood}. In addition, heterogeneous information networks, where users are represented as nodes and various relationships serve as links, are also employed to detect social bots \cite{peng2024unsupervised,feng2022heterogeneity,li2022sybilflyover}.

Although heterogeneous information networks, which represent various relationships as network links, are widely applied, the heterogeneity of neighborhood preferences is often overlooked. Additionally, unsupervised methods such as SybilRank, SybilWalk, and SybilSCAR are typically defined empirically, and GNNs frequently lack interpretability or intuitive design. To address these challenges, we propose a systematic, theoretical, and intuitive framework for social bot detection based on the Na\"ive Bayes model. This framework emphasizes heterogeneous motifs derived exclusively from network topology, incorporating neighborhood preferences to enhance detection capabilities.

\section{Preliminaries}
\label{section4}

\subsection{Na\"ive Bayes model}
\label{section41}
The Na\"ive Bayes model is a probabilistic and solid mathematical algorithm based on the Bayesian theorem, which assumes all features are conditionally independent given the target label \cite{bernardo2009bayesian}. This allows the model to calculate the probability of a target efficiently by treating the joint likelihood of features as the product of their probabilities. Despite its simplicity, the Na\"ive Bayes model performs well in various applications and offers the advantages of scalability, computational efficiency, and interpretability.

Let $Y$ be a label variable and $X=\{x_1, x_2, ..., x_n\}$ denote its feature vector, meaning the feature vector $X$ is associated with the label $Y$. According to Bayesian theory, the probability $P(Y|\{x_1, x_2, ..., x_n\})$ can be given as
\begin{align}
P(Y|\{x_1, x_2, ..., x_n\}) &= \frac{P(Y)P(\{x_1, x_2, ..., x_n\}|Y)}{P(\{x_1, x_2, ..., x_n\})},
\label{equation:nbm}
\end{align}
where $P(Y)$ is the prior probability of $Y$, $P(X|Y)$ is the likelihood of $X$ relative to $Y$.

In Na\"ive Bayes model, the assumption of conditional independence allows the conditional probability $P(\{x_1, x_2, ..., x_n\}|Y)$ can be expressed as
\begin{align}
P(\{x_1, x_2, ..., x_n\}|Y) &= \prod_{i=1}^{n} P(x_i|Y).
\end{align}
Hence, Eq. (\ref{equation:nbm}) can be rewritten as
\begin{align}
P(Y|\{x_1, x_2, ..., x_n\}) &= \frac{P(Y)\prod_{i=1}^{n} P(x_i|Y)}{P(\{x_1, x_2, ..., x_n\})}.
\end{align}

\subsection{Performance metrics}
\label{section42}
In social bot detection, standard evaluation metrics include Accuracy, Precision, Recall, F1 score, and the area under the receiver operating characteristic curve (AUC) \cite{yang2020scalable,yang2022botometer,dehghan2023detecting,huang2024cgnn,peng2024unsupervised}. Accuracy measures the overall proportion of correct predictions but may be misleading in imbalanced datasets. Precision reflects the proportion of true bots among all accounts predicted as bots, while Recall measures the proportion of actual bots correctly identified. Each metric has limitations when used independently.

To address this, the F1 score---the harmonic mean of Precision and Recall---provides a balanced measure of classification performance. AUC evaluates a model's ability to distinguish between the positive test set (bots set) $T^P$ and the negative test set (humans set) $T^N$ by comparing predicted probabilities. It is particularly useful in imbalanced settings, offering a more comprehensive assessment of classifier effectiveness. Together, these metrics provide a robust evaluation framework for social bot detection.

\subsection{Feature analyses}
\label{section43}
Feature selection techniques aim to improve model performance and reduce complexity by identifying the most relevant features. They are generally categorized into filter, wrapper, and embedded methods \cite{li2017feature}. Filter methods (\textit{e.g.}, Pearson and Spearman correlation, maximum information coefficient) evaluate features independently of any model and are computationally efficient but may miss feature interactions. Wrapper methods assess feature subsets by training models on them, offering higher accuracy at the cost of increased computation and potential overfitting. Embedded methods perform feature selection during model training, balancing efficiency and performance, but are model-dependent.

To overcome these weaknesses, we consider a theoretical framework to quantify the maximum capability of each feature in bot detection \cite{ran2024maximum}. The theoretical framework is independent of any machine learning-based model and can mathematically quantify the maximum capability of each feature. A feature is proposed to represent the social relationships and neighborhood preferences within the target node's environment. Therefore, target nodes that do not contain the feature are assigned the value 0. Assume that $T_1$ is the subset of $T^P$ in which target nodes hold the feature, while the complement set $\overline{T}_1$ is composed of target nodes that do not hold the feature. Similarly, the negative set $T^N$ can also be divided into two subsets $T_2$ and $\overline{T}_2$. Assume that $T_1$ takes a fraction $p_1$ of $T^P$ and $T_2$ takes a fraction $p_2$ of $T^N$. Because all target nodes in $\overline{T}_1$ and $\overline{T}_2$ have the same value 0, the detection performance mainly relies on the ranking of $T_1$ and $T_2$.

According to the mathematical expression derived by Ran \textit{et al.} \cite{ran2024maximum}, the AUC upper bound of a feature by a machine learning-based classifier in bot detection can be calculated as
\begin{align}
\text{AUC}^{\prime}_\text{upper} &= \frac{n'}{n} + \frac{1}{2}\frac{n''}{n} \nonumber \\
            &= p_{1} + (1-p_{1})p_{2} +\frac{1}{2}(1-p_{1})(1-p_{2}) \nonumber \\
            &=  \frac{1}{2}+\frac{p_{1}+p_{2}-p_{1}p_{2}}{2}. \label{equation:upper2}                                   
\end{align} 
Eq. (\ref{equation:upper2}) effectively provides the maximum capability of a feature in supervised bot detection.

\section{Methodology}
\label{section3}

Complex systems are often represented as networks, where nodes represent components of the system and links represent interactions between them \cite{allen2024quantifying}. These systems can typically be modeled as heterogeneous networks with different types of nodes and links \cite{peng2024unsupervised,feng2022heterogeneity,li2022sybilflyover,shi2018heterogeneous}. In this work, we focus on real network structures derived from the following (or follower) relationships and construct heterogeneous networks by incorporating node-label information.

Identifying and analyzing network motifs can uncover the underlying principles and mechanisms governing various networks, such as biological, social, and technological systems. Motifs have been successfully applied to a myriad of problems such as link prediction \cite{liu2019link}, graph classification \cite{peng2020motif}, protein function detection \cite{jones2003using}, and molecular property prediction \cite{zhang2021motif}.

\subsection{Homogeneous motifs}
\label{section31}

\begin{figure}[htbp]
	\centering
	\includegraphics[width=0.48\textwidth]{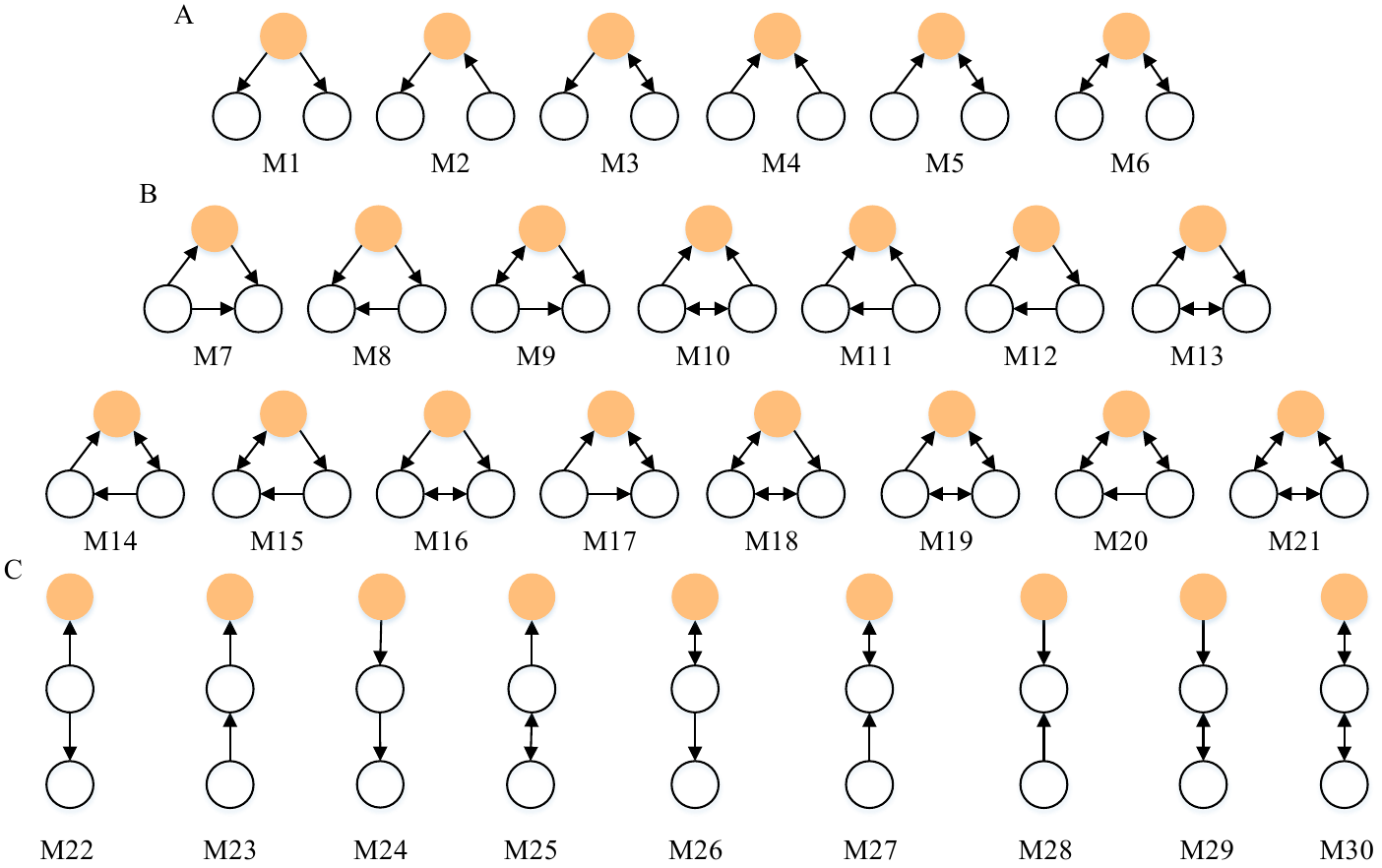}
	\caption{The 30 unique 3-node homogeneous motifs in a directed network. The yellow node is the target node to be detected. A) 6 first-order homogeneous motifs. B) 15 closed homogeneous motifs. C) 9 second-order homogeneous motifs.}
	\label{Fig:hm}
\end{figure}

Here, we examine directed homogeneous networks, which can give rise to homogeneous motifs. Since the network is directed, we can develop 30 unique 3-node homogeneous motifs based on the different positions of the target node to be detected. In this paper, the homogeneous motifs we consider differ from traditional ones. Traditional homogeneous motifs treat local structures as a whole, without accounting for the varying positions of the target node \cite{milo2002network,alon2007network}. For instance, M3 and M25 in Fig. \ref{Fig:hm} would be considered identical for traditional homogeneous motifs. Here, two factors focus exclusively on motifs with fewer than three nodes. First, calculating these smaller motifs involves lower computational complexity. Second, higher-order motifs are typically made up of lower-order ones, and their abundance depends on the prevalence of these simpler structures \cite{vazquez2004topological}.

In this work, we assume that the yellow node represents the target node to be detected. The 30 unique 3-node homogeneous motifs can be categorized into three types (Fig. \ref{Fig:hm}). First-order motifs are constructed by starting from the target node and identifying its immediate neighbors. Closed motifs are formed by starting from the target node and identifying two adjacent nodes that are also connected. Second-order motifs are created by starting from the target node, identifying its neighbors, and then finding the neighbors of these neighbors.

\subsection{Heterogeneous motifs}
\label{section32}
\begin{figure}[htbp]
	\centering
	\includegraphics[width=0.48\textwidth]{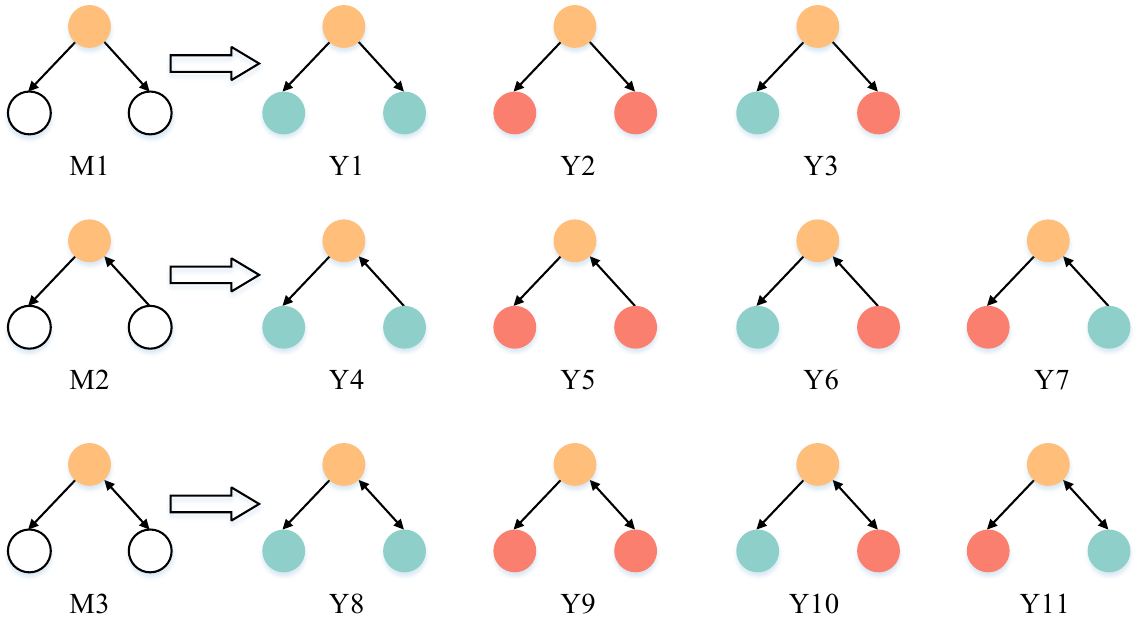}
	\caption{The heterogeneous motifs by refining homogeneous motifs with node-label information. The yellow node is the target node to be detected. The green and red nodes are human and bot, respectively.}
	\label{Fig:hm2hm}
\end{figure}

To capture the heterogeneity of neighborhood preferences, we propose heterogeneous motifs by refining homogeneous motifs with the incorporation of node-label information. By utilizing different labels of neighboring nodes around the target node, we can construct a variety of structurally diverse heterogeneous motifs. For example, we can refine the homogeneous motif M1 in Fig. \ref{Fig:hm} into three heterogeneous motifs (Y1, Y2, and Y3 in Fig. \ref{Fig:hm2hm}). By subdividing each homogeneous motif in Fig. \ref{Fig:hm}, we yield 114 distinctive 3-node heterogeneous motifs. These refined heterogeneous motifs effectively capture the social relationships and neighborhood preferences within the target node's environment. 

\subsection{Theoretical framework based on Na\"ive Bayes model}
\label{section33}
\begin{figure}[htbp]
	\centering
	\resizebox{5cm}{!}{\includegraphics{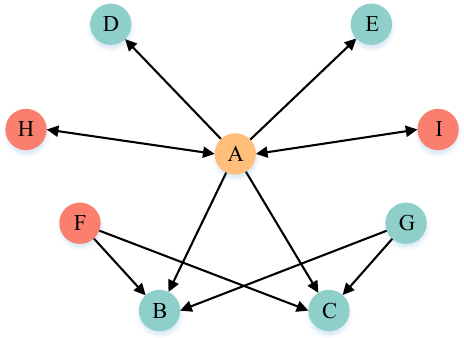}}
	\caption{An example demonstrating the differing contributions of node pairs $(B, C)$ and $(D, E)$ to the probability of $A$ being identified as a social bot. The node $A$ is the target node under evaluation. For heterogeneous motif Y1, the node pairs $(B, C)$ along with nodes $F$ and $G$ can successfully form its motif Y1. In contrast, the node pairs $(D, E)$ are unable to form motif Y1, highlighting their differing roles in the detection process.}
	\label{fig:nbm}
\end{figure}

Current topology-based social bot detection methods, such as Ising, SybilWalk, and SybilSCAR, are primarily intuitive and experience-driven, lacking a systematic theoretical framework to quantify the contribution of different node pairs. Here, we describe and quantify these differences. To determine whether node $A$ is a social bot, we can consider the neighborhood information provided by nodes $B$, $C$, $D$, $E$, $H$, and $I$ (Fig. \ref{fig:nbm}). Using the heterogeneous motif Y1 as an example, we can calculate the probability of node $A$ being a social bot by leveraging the motif Y1 formed by node pairs $(B, C)$, $(B, D)$, $(C, E)$, and $(D, E)$ with node A. Since the pair $(B, C)$ forms the motif Y1 with nodes $F$ and $G$, the pairs $(B, D)$, $(C, E)$, and $(D, E)$ do not, leading to differing contributions from these two pairs to the probability of $A$ being identified as a social bot. In the following, we use Bayesian theory to quantify these differences systematically.

For a given directed network $G(N, L, W)$, where $N$ is the set of nodes, $L$ is the set of links, and $W$ is the set of labels associated with the nodes in $N$. A node in network $G$ cannot connect to itself (no self-loops) nor share more than one link with another node (no repeated links). Let $A_{h}$ and $A_{b}$ denote a human and a bot for the node $A$, respectively. Here, we assume that the node $A$ needs to be detected. Based on Bayesian theory, the posterior probability of $A$ being human or bot can be calculated as
\begin{equation}
P(A_{b}|S_\text{i}(A)) = \frac{P(A_{b})P(S_\text{i}(A)|A_{b})}{P(S_\text{i}(A))},
\label{equation3}
\end{equation}
\begin{equation}
P(A_{h}|S_\text{i}(A)) = \frac{P(A_{h})P(S_\text{i}(A)|A_{h})}{P(S_\text{i}(A))},
\label{equation4}
\end{equation}
where $S_\text{i}(A)$ denotes a set of node pairs that form a homogeneous motif Mi (heterogeneous motif Yi) in conjunction with node $A$.

For the target node $A$, the label of node $A$ is determined using Maximum A Posteriori Estimation \cite{greig1989exact} by computing the ratio of these two posterior probabilities. Assume that the contribution of each node pair $(B, C)$ in $S_\text{i}(A)$ to the probability of node $A$ being a human or bot is independent. Hence, the likelihood score for the label of target node $A$ can be calculated as
\begin{align}
r_{A}  &= \frac{P(A_{b}|S_\text{i}(A))}{P(A_{h}|S_\text{i}(A))} = \frac{P(A_{b})}{P(A_{h})} \frac{P(S_\text{i}(A)|A_{b})}{P(S_\text{i}(A)|A_{h})} \nonumber \\
            &= \frac{P(A_{b})}{P(A_{h})} \prod_{(B,C) \in S_\text{i}(A)} \frac{P((B,C)|A_{b})}{P((B,C)|A_{h})} \nonumber \\
            &= \frac{P(A_{b})}{P(A_{h})} \prod_{(B,C) \in S_\text{i}(A)} \frac{P(A_{h})}{P(A_{b})} \frac{P(A_{b}|(B,C))}{P(A_{h}|(B,C))}.
\label{equation5}
\end{align}
Here, $P(A_{b})$ represents the prior probability that node $A$ is a bot, and $P(A_{h})$ denotes the prior probability that node $A$ is a human. Hence, $P(A_{b})$ and $P(A_{h})$ can be computed as
\begin{equation}
P(A_{b}) = \frac{p|N|}{|N|}=f_b,
\label{equation6}
\end{equation}
\begin{equation}
P(A_{h}) = \frac{n|N|}{|N|}=f_h,
\label{equation7}
\end{equation}
where $f_b$ and $f_h$ are the fraction of bots and humans in a network, respectively. Let $a = \frac{P(A_{h})}{P(A_{b})}=\frac{f_h}{f_b}$, Eq. (\ref{equation5}) can be rewritten as
\begin{equation}
r_{A} = a^{-1} \prod_{(B,C) \in S_\text{i}(A)}a {R_{S_\text{i}((B,C))}},
\label{equation8}
\end{equation}
where $R_{S_\text{i}((B,C))}= \frac{P(A_{b}|(B,C))}{P(A_{h}|(B,C))}$. Since $a$ is a constant for a given network, the factor $a^{-1}$ can be omitted. Taking the logarithm of Eq. (\ref{equation8}), the new likelihood score for the label of target node $A$ can be calculated as
\begin{align}
\tilde{r}_{A} &= \log( \prod_{(B,C) \in S_\text{i}(A)}a {R_{S_\text{i}((B,C))}}) \nonumber \\
                     &= \sum_{(B,C) \in S_\text{i}(A)}\log(a{R_{S_\text{i}((B,C))}}) \nonumber \\
                     &= |S_\text{i}(A)|\log{a}  + \sum_{(B,C) \in S_\text{i}(A)}\log{R_{S_\text{i}((B,C))}},
\label{equation9}
\end{align}
where $|S_\text{i}(A)|$ denotes the number of the set of node pairs that form a homogeneous motif Mi (heterogeneous motif Yi) in conjunction with node $A$.

By defining the role function $R_{S_\text{i}((B,C))}$, $R_{S_\text{i}((B,C))}$ in Eq. (\ref{equation9}) can be estimated as
\begin{align}
R_{S_\text{i}((B,C))} &= \frac{P(A_{b}|(B,C))}{P(A_{h}|(B,C))}= \frac{\frac{N_{b_{S_\text{i}((B,C))}}}{N_{b_{S_\text{i}((B,C))}} + N_{h_{S_\text{i}((B,C))}}}}{\frac{N_{h_{S_\text{i}((B,C))}}}{N_{b_{S_\text{i}((B,C))}} + N_{h_{S_\text{i}((B,C))}}}} \nonumber \\
            &= \frac{N_{b_{S_\text{i}((B,C))}}}{N_{h_{S_\text{i}((B,C))}}},
\label{equation10}
\end{align}
where $N_{b_{S_\text{i}((B,C))}}$ represents the number of motifs $S_\text{i}$ formed by a bot node with nodes $B$ and $C$. Similarly, $N_{h_{S_\text{i}((B,C))}}$ denotes the number of motifs $S_\text{i}$ formed by a human node with nodes $B$ and $C$. To avoid the expression becoming meaningless due to a zero denominator, 1 is added to both the numerator and denominator in Eq. (\ref{equation10}). Hence, Eq. (\ref{equation10}) can be defined as
\begin{equation}
\tilde{R}_{S_\text{i}((B,C))} = \frac{N_{b_{S_\text{i}((B,C))}}+1}{N_{h_{S_\text{i}((B,C))}}+1}.
\label{equation11}
\end{equation}
By applying Eq. (\ref{equation11}) to Eq. (\ref{equation9}), we can quantify the likelihood values for each homogeneous and heterogeneous motif in the context of social bot detection.

\subsection{Time complexity analysis}
\label{section34}
Discovering network motifs is a computationally challenging task, as it involves matching target motifs with subgraph patterns in a network. Here, we employ G-Tries, an efficient data structure designed for motif discovery \cite{ribeiro2010g}. G-Tries constructs a tree-based structure, known as a g-trie, by inserting a collection of networks, each representing a unique subgraph pattern. The complexity of this construction is primarily determined by two steps: recursive insertion and canonical labeling.

The recursive insertion procedure is invoked for each network inserted into the g-trie. This process operates over the adjacency matrix of the input network, resulting in a time complexity of $O(k^2)$, where $k$ is the number of nodes in the motif.

Additionally, each graph must undergo canonical labeling to ensure structural uniqueness within the g-trie. This step is computationally intensive, as it involves solving the graph isomorphism problem. In the worst case, it requires $O(k!)$ time due to the $k!$ possible permutations of node labels.

Assuming $N$ motifs of size $k$ are inserted, the overall worst-case time complexity for constructing the g-trie is $O(N*k!)$. In this work, we focus exclusively on 3-node motifs. Therefore, the time complexity for processing homogeneous motifs is $O(30*3!)$, while that for heterogeneous motifs is $O(114*3!)$.

\subsection{XGBoost classifier}
\label{section35}
XGBoost has gained wide recognition as one of the most powerful algorithms for classification tasks \cite{chen2016xgboost}. Its most notable advantage lies in its high predictive accuracy. By sequentially constructing an ensemble of decision trees in which each tree corrects the errors of its predecessors, XGBoost effectively captures complex non-linear patterns in data. This capability makes it particularly suitable for structured datasets where relationships among features are subtle and often overlooked by simpler models.

Another important strength of XGBoost is its robust regularization framework, which incorporates both L1 and L2 penalties to control model complexity. These mechanisms help mitigate overfitting, a common challenge in boosting algorithms, and ensure that the model generalizes well to unseen data. Furthermore, XGBoost offers a rich set of parameters, such as maximum tree depth, minimum child weight, and learning rate, which provide fine-grained control over the trade-off between complexity and predictive performance.

XGBoost has also been widely applied in social bot detection, where it leverages heterogeneous user features, including profile metadata, behavioral patterns, and network structures, to distinguish between bots and human users \cite{tang2025uncovering}. Beyond strong predictive performance, XGBoost supports interpretability and flexibility. It provides built-in measures of feature importance, enabling researchers to identify which attributes contribute most to classification decisions. Its flexibility further extends to accommodating diverse objectives and evaluation metrics, making it well suited for a broad range of classification problems.

In this paper, we use the likelihood score computed by Eq. (\ref{equation9}) as either a one-dimensional feature or a set of multiple likelihood scores as multi-dimensional features, which are then fed into an XGBoost classifier to perform social bot detection.

\section{Experiments and Results}
\label{section5}

\subsection{Dataset preprocessing}
\label{section51}
Here, we consider four publicly available datasets: Cresci-15 \cite{cresci2015fame}, MGTAB \cite{shi2025mgtab}, TwiBot-20 \cite{feng2021twibot}, and TwiBot-22 \cite{feng2022twibot} to evaluate the detection performance of homogeneous and heterogeneous motifs by the theoretical framework based on Na\"ive Bayes model.

In this study, we focus on follower and following relationships to develop real homogeneous and heterogeneous motifs from real network structures. After filtering the original dataset, the statistics of all datasets we deal with are shown in Table~\ref{tab:table1}.

\begin{table}[htbp]
\caption{Statistics of social bot detection datasets. $|N|$ and $|L|$ denote the number of nodes and links, respectively. $\#$ Human represents the number of humans. $\#$ Bot refers to the number of social bots.}
\resizebox{\columnwidth}{!}{
\begin{tabular}{cccccc}
\hline
Datasets & $|N|$ & $|L|$ & $\#$ Human & $\#$ Bot \\
\hline
Cresci-15 & 1,741& 6,214 & 1,105 & 636 \\
MGTAB & 9,443 & 425,863 & 6,968 & 2,475 \\
TwiBot-20 & 205,730 & 227,477 & 5,237 & 6,589\\
TwiBot-22 & 693,761 & 3,711,903 & 612,329 & 81,432 \\
\hline
\end{tabular}
}
\label{tab:table1}
\end{table}

\subsection{Social bot detection baselines}
\label{section53}
To evaluate the effectiveness of the proposed homogeneous and heterogeneous motifs, we compare our approach against a broad range of baseline methods spanning both unsupervised and supervised learning paradigms. The unsupervised baselines---Ising \cite{des2022detecting}, SybilWalk \cite{jia2017random}, and SybilSCAR \cite{wang2018structure}---primarily leverage graph structures and label propagation techniques to detect social bots. In contrast, Botometer \cite{yang2022botometer}, FP \cite{moghaddam2022friendship}, and ARG \cite{abreu2020twitter} represent feature-based supervised learning methods that utilize user metadata, friendship patterns, and publicly available attributes for classification. More advanced deep learning models, including DeeProBot \cite{hayawi2022deeprobot}, BotRGCN \cite{feng2021botrgcn}, T5 \cite{raffel2020exploring}, and RGT \cite{feng2022heterogeneity}, incorporate textual content, multi-modal features, or heterogeneous network structures through architectures such as LSTMs, graph convolutional networks, and transformers. This diverse set of baselines enables a comprehensive evaluation of detection strategies, ranging from structural heuristics to representation learning.

In this study, social bot detection is formulated as a binary classification task. We adopt an XGBoost classifier and perform hyperparameter tuning to optimize model performance. To ensure reliable and robust evaluation, we apply ten-fold cross-validation \cite{yang2020scalable}, maintaining a balanced ratio of positive and negative samples in each fold. A grid search procedure \cite{bergstra2012random} is used to identify the optimal hyperparameter configuration that maximizes the AUC.

\subsection{Results and analysis}
\label{section54}

\subsubsection{Results on homogeneous motifs}
To evaluate the effectiveness of each homogeneous motif, we use the likelihood score computed by Eq. (\ref{equation9}) as a one-dimensional feature and feed it into an XGBoost classifier to perform social bot detection. As shown in Table \ref{table:homo}, second-order homogeneous motifs tend to outperform both first-order and closed motifs across all four datasets. Notably, the second-order motif M30 demonstrates strong performance, particularly on the Cresci-15 dataset. This performance gap can be attributed to differences in structural complexity: first-order motifs encode only basic dyadic interactions, which provide limited structural information and are vulnerable to noise or simple evasion strategies used by bots. In contrast, second-order motifs capture richer local structures---such as triadic closures and multi-link patterns---that reflect more complex social dynamics and are harder for bots to mimic. These properties make second-order motifs more discriminative and generalizable across diverse user communities.

Moreover, differences in performance between motifs such as M3 and M25 further support our hypothesis that the position of the target node within a motif significantly influences detection outcomes (Table \ref{table:homo}). Finally, the combined use of all 30 homogeneous motifs, denoted as Homogeneous motifs (All), achieves the best overall performance across all datasets and metrics (Table \ref{table:homo}). This demonstrates the complementary strengths of different motif types: while individual motifs highlight specific structural patterns, their combination provides a more comprehensive view of local topology. By integrating first-order, closed, and second-order motifs, the classifier leverages both simple and complex structural cues, significantly enhancing its ability to distinguish bots from real users. The consistent and robust performance of Homogeneous motifs (All) underscores the value of structural diversity in social bot detection.

\begin{table*}[htbp]
	\centering
	\caption{Performance comparison of the average Accuracy, Precision, Recall, and F1 achieved by homogeneous motifs using an XGBoost classifier for social bot detection. The overall optimal results under the four metrics are marked in bold.}
        \resizebox{\textwidth}{!}{
	\begin{tabular}{l|rrrrrrrrrrrrrrrr}
	\hline
	\multicolumn{1}{c}{\multirow{2}*{Features}}& \multicolumn{4}{c|}{Cresci-15}& \multicolumn{4}{c|}{MGTAB}& \multicolumn{4}{c|}{TwiBot-20}& \multicolumn{4}{c}{TwiBot-22}\\
	\cline{2-17}
	\multicolumn{1}{c}{}& \multicolumn{1}{c}{Accuracy} & \multicolumn{1}{c}{Precision} & \multicolumn{1}{c}{Recall}& \multicolumn{1}{c|}{F1}& \multicolumn{1}{c}{Accuracy} & \multicolumn{1}{c}{Precision} & \multicolumn{1}{c}{Recall}& \multicolumn{1}{c|}{F1}&\multicolumn{1}{c}{Accuracy} & \multicolumn{1}{c}{Precision} & \multicolumn{1}{c}{Recall}& \multicolumn{1}{c|}{F1}&\multicolumn{1}{c}{Accuracy} & \multicolumn{1}{c}{Precision} & \multicolumn{1}{c}{Recall}& \multicolumn{1}{c}{F1}\\
		  \hline		
	  \multicolumn{1}{c}{M1} & \multicolumn{1}{c}{0.497} & \multicolumn{1}{c}{0.198}& \multicolumn{1}{c}{0.400}& \multicolumn{1}{c|}{0.265} & \multicolumn{1}{c}{0.730}& \multicolumn{1}{c}{0.666} & \multicolumn{1}{c}{0.922}& \multicolumn{1}{c|}{0.773}&\multicolumn{1}{c}{0.648}& \multicolumn{1}{c}{0.640} & \multicolumn{1}{c}{0.924}& \multicolumn{1}{c|}{0.756}&\multicolumn{1}{c}{0.579}& \multicolumn{1}{c}{0.545} & \multicolumn{1}{c}{0.833}& \multicolumn{1}{c}{0.659} \\
      
	\multicolumn{1}{c}{M2} & \multicolumn{1}{c}{0.497} & \multicolumn{1}{c}{0.198}& \multicolumn{1}{c}{0.400}& \multicolumn{1}{c|}{0.265} & \multicolumn{1}{c}{0.701}& \multicolumn{1}{c}{0.672} & \multicolumn{1}{c}{0.787}& \multicolumn{1}{c|}{0.725}&\multicolumn{1}{c}{0.650}& \multicolumn{1}{c}{0.646} & \multicolumn{1}{c}{0.903}& \multicolumn{1}{c|}{0.753}& \multicolumn{1}{c}{0.637}&\multicolumn{1}{c}{0.582} & \multicolumn{1}{c}{0.907}& \multicolumn{1}{c}{0.709} \\
    
	\multicolumn{1}{c}{M3} & \multicolumn{1}{c}{0.497} & \multicolumn{1}{c}{0.198}& \multicolumn{1}{c}{0.400}& \multicolumn{1}{c|}{0.265} & \multicolumn{1}{c}{0.732}& \multicolumn{1}{c}{0.670} & \multicolumn{1}{c}{0.916}& \multicolumn{1}{c|}{0.774}&\multicolumn{1}{c}{0.590}& \multicolumn{1}{c}{0.590} & \multicolumn{1}{c}{0.999}& \multicolumn{1}{c|}{0.742}& \multicolumn{1}{c}{0.617}&\multicolumn{1}{c}{0.567} & \multicolumn{1}{c}{0.902}& \multicolumn{1}{c}{0.696} \\
    
	\multicolumn{1}{c}{M4} &\multicolumn{1}{c}{0.497} & \multicolumn{1}{c}{0.198}& \multicolumn{1}{c}{0.400}& \multicolumn{1}{c|}{0.265} & \multicolumn{1}{c}{0.580}&\multicolumn{1}{c}{0.562} & \multicolumn{1}{c}{0.727}& \multicolumn{1}{c|}{0.634}&\multicolumn{1}{c}{0.631}&\multicolumn{1}{c}{0.622} & \multicolumn{1}{c}{0.955}& \multicolumn{1}{c|}{0.753}& \multicolumn{1}{c}{0.765}&\multicolumn{1}{c}{0.709} & \multicolumn{1}{c}{0.880}& \multicolumn{1}{c}{0.785} \\
    
	\multicolumn{1}{c}{M5} &\multicolumn{1}{c}{0.497} & \multicolumn{1}{c}{0.198}& \multicolumn{1}{c}{0.400}& \multicolumn{1}{c|}{0.265} & \multicolumn{1}{c}{0.671}&\multicolumn{1}{c}{0.654} & \multicolumn{1}{c}{0.728}& \multicolumn{1}{c|}{0.689}&\multicolumn{1}{c}{0.590}&\multicolumn{1}{c}{0.590} & \multicolumn{1}{c}{0.998}& \multicolumn{1}{c|}{0.742}& \multicolumn{1}{c}{0.641}&\multicolumn{1}{c}{0.585} & \multicolumn{1}{c}{0.918}& \multicolumn{1}{c}{0.714} \\
    
	\multicolumn{1}{c}{M6} &\multicolumn{1}{c}{0.721} & \multicolumn{1}{c}{0.675}& \multicolumn{1}{c}{0.855}& \multicolumn{1}{c|}{0.754} & \multicolumn{1}{c}{0.708}&\multicolumn{1}{c}{0.660} & \multicolumn{1}{c}{0.856}& \multicolumn{1}{c|}{0.745}&\multicolumn{1}{c}{0.590}&\multicolumn{1}{c}{0.590} & \multicolumn{1}{c}{0.999}& \multicolumn{1}{c|}{0.743}& \multicolumn{1}{c}{0.608}&\multicolumn{1}{c}{0.559} & \multicolumn{1}{c}{0.940}& \multicolumn{1}{c}{0.701} \\
    
	\multicolumn{1}{c}{M7} &\multicolumn{1}{c}{0.497} & \multicolumn{1}{c}{0.198}& \multicolumn{1}{c}{0.400}& \multicolumn{1}{c|}{0.266} & \multicolumn{1}{c}{0.690}&\multicolumn{1}{c}{0.646} & \multicolumn{1}{c}{0.841}& \multicolumn{1}{c|}{0.730}&\multicolumn{1}{c}{0.591}&\multicolumn{1}{c}{0.591} & \multicolumn{1}{c}{0.999}& \multicolumn{1}{c|}{0.743}& \multicolumn{1}{c}{0.562}&\multicolumn{1}{c}{0.528} & \multicolumn{1}{c}{0.971}& \multicolumn{1}{c}{0.684} \\
    
	\multicolumn{1}{c}{M8} &\multicolumn{1}{c}{0.497} & \multicolumn{1}{c}{0.198}& \multicolumn{1}{c}{0.400}& \multicolumn{1}{c|}{0.265} & \multicolumn{1}{c}{0.689}&\multicolumn{1}{c}{0.622} & \multicolumn{1}{c}{0.965}& \multicolumn{1}{c|}{0.756}&\multicolumn{1}{c}{0.589}&\multicolumn{1}{c}{0.590} & \multicolumn{1}{c}{0.998}& \multicolumn{1}{c|}{0.742}& \multicolumn{1}{c}{0.544}&\multicolumn{1}{c}{0.518} & \multicolumn{1}{c}{0.924}& \multicolumn{1}{c}{0.664} \\
    
	\multicolumn{1}{c}{M9} &\multicolumn{1}{c}{0.497} & \multicolumn{1}{c}{0.198}& \multicolumn{1}{c}{0.400}& \multicolumn{1}{c|}{0.265} & \multicolumn{1}{c}{0.713}&\multicolumn{1}{c}{0.648} & \multicolumn{1}{c}{0.929}& \multicolumn{1}{c|}{0.764}&\multicolumn{1}{c}{0.590}&\multicolumn{1}{c}{0.590} & \multicolumn{1}{c}{0.998}& \multicolumn{1}{c|}{0.742}& \multicolumn{1}{c}{0.547}&\multicolumn{1}{c}{0.519} & \multicolumn{1}{c}{0.970}& \multicolumn{1}{c}{0.676} \\
    
	\multicolumn{1}{c}{M10} &\multicolumn{1}{c}{0.497} & \multicolumn{1}{c}{0.198}& \multicolumn{1}{c}{0.400}& \multicolumn{1}{c|}{0.265} & \multicolumn{1}{c}{0.551}&\multicolumn{1}{c}{0.535} & \multicolumn{1}{c}{0.776}& \multicolumn{1}{c|}{0.633}&\multicolumn{1}{c}{0.590}&\multicolumn{1}{c}{0.590} & \multicolumn{1}{c}{0.999}& \multicolumn{1}{c|}{0.743}& \multicolumn{1}{c}{0.610}&\multicolumn{1}{c}{0.558} & \multicolumn{1}{c}{0.971}& \multicolumn{1}{c}{0.709} \\
    
	\multicolumn{1}{c}{M11} &\multicolumn{1}{c}{0.497} & \multicolumn{1}{c}{0.198}& \multicolumn{1}{c}{0.400}& \multicolumn{1}{c|}{0.265} & \multicolumn{1}{c}{0.607}&\multicolumn{1}{c}{0.579} & \multicolumn{1}{c}{0.784}& \multicolumn{1}{c|}{0.667}&\multicolumn{1}{c}{0.591}&\multicolumn{1}{c}{0.591} & \multicolumn{1}{c}{0.999}& \multicolumn{1}{c|}{0.742}& \multicolumn{1}{c}{0.691}&\multicolumn{1}{c}{0.621} & \multicolumn{1}{c}{0.940}& \multicolumn{1}{c}{0.748} \\
    
	\multicolumn{1}{c}{M12} &\multicolumn{1}{c}{0.497} & \multicolumn{1}{c}{0.198}& \multicolumn{1}{c}{0.400}& \multicolumn{1}{c|}{0.265} & \multicolumn{1}{c}{0.647}&\multicolumn{1}{c}{0.593} & \multicolumn{1}{c}{0.937}& \multicolumn{1}{c|}{0.727}&\multicolumn{1}{c}{0.590}&\multicolumn{1}{c}{0.590} & \multicolumn{1}{c}{0.999}& \multicolumn{1}{c|}{0.743}& \multicolumn{1}{c}{0.553}&\multicolumn{1}{c}{0.523} & \multicolumn{1}{c}{0.980}& \multicolumn{1}{c}{0.682} \\
    
	\multicolumn{1}{c}{M13} &\multicolumn{1}{c}{0.497} & \multicolumn{1}{c}{0.198}& \multicolumn{1}{c}{0.400}& \multicolumn{1}{c|}{0.265} & \multicolumn{1}{c}{0.687}&\multicolumn{1}{c}{0.630} & \multicolumn{1}{c}{0.897}& \multicolumn{1}{c|}{0.740}&\multicolumn{1}{c}{0.590}&\multicolumn{1}{c}{0.590} & \multicolumn{1}{c}{0.999}& \multicolumn{1}{c|}{0.743}& \multicolumn{1}{c}{0.545}&\multicolumn{1}{c}{0.518} & \multicolumn{1}{c}{0.981}& \multicolumn{1}{c}{0.678} \\
    
	\multicolumn{1}{c}{M14} &\multicolumn{1}{c}{0.497} & \multicolumn{1}{c}{0.198}& \multicolumn{1}{c}{0.400}& \multicolumn{1}{c|}{0.265} & \multicolumn{1}{c}{0.649}&\multicolumn{1}{c}{0.618} & \multicolumn{1}{c}{0.781}& \multicolumn{1}{c|}{0.690}&\multicolumn{1}{c}{0.590}&\multicolumn{1}{c}{0.590} & \multicolumn{1}{c}{0.999}& \multicolumn{1}{c|}{0.743}& \multicolumn{1}{c}{0.558}&\multicolumn{1}{c}{0.525} & \multicolumn{1}{c}{0.980}& \multicolumn{1}{c}{0.684} \\

    \multicolumn{1}{c}{M15} &\multicolumn{1}{c}{0.497} & \multicolumn{1}{c}{0.198}& \multicolumn{1}{c}{0.400}& \multicolumn{1}{c|}{0.265} & \multicolumn{1}{c}{0.681}&\multicolumn{1}{c}{0.615} & \multicolumn{1}{c}{0.964}& \multicolumn{1}{c|}{0.751}&\multicolumn{1}{c}{0.590}&\multicolumn{1}{c}{0.590} & \multicolumn{1}{c}{0.999}& \multicolumn{1}{c|}{0.742}& \multicolumn{1}{c}{0.536}&\multicolumn{1}{c}{0.513} & \multicolumn{1}{c}{0.981}& \multicolumn{1}{c}{0.673} \\

    \multicolumn{1}{c}{M16} &\multicolumn{1}{c}{0.407} & \multicolumn{1}{c}{0.198}& \multicolumn{1}{c}{0.400}& \multicolumn{1}{c|}{0.265} & \multicolumn{1}{c}{0.681}&\multicolumn{1}{c}{0.617} & \multicolumn{1}{c}{0.954}& \multicolumn{1}{c|}{0.750}&\multicolumn{1}{c}{0.591}&\multicolumn{1}{c}{0.591} & \multicolumn{1}{c}{0.999}& \multicolumn{1}{c|}{0.743}& \multicolumn{1}{c}{0.531}&\multicolumn{1}{c}{0.511} & \multicolumn{1}{c}{0.944}& \multicolumn{1}{c}{0.663} \\

    \multicolumn{1}{c}{M17} &\multicolumn{1}{c}{0.497} & \multicolumn{1}{c}{0.198}& \multicolumn{1}{c}{0.400}& \multicolumn{1}{c|}{0.265} & \multicolumn{1}{c}{0.663}&\multicolumn{1}{c}{0.633} & \multicolumn{1}{c}{0.777}& \multicolumn{1}{c|}{0.698}&\multicolumn{1}{c}{0.590}&\multicolumn{1}{c}{0.590} & \multicolumn{1}{c}{0.999}& \multicolumn{1}{c|}{0.743}& \multicolumn{1}{c}{0.552}&\multicolumn{1}{c}{0.522} & \multicolumn{1}{c}{0.978}& \multicolumn{1}{c}{0.681} \\

    \multicolumn{1}{c}{M18} &\multicolumn{1}{c}{0.497} & \multicolumn{1}{c}{0.198}& \multicolumn{1}{c}{0.400}& \multicolumn{1}{c|}{0.265} & \multicolumn{1}{c}{0.701}&\multicolumn{1}{c}{0.643} & \multicolumn{1}{c}{0.905}& \multicolumn{1}{c|}{0.752}&\multicolumn{1}{c}{0.590}&\multicolumn{1}{c}{0.590} & \multicolumn{1}{c}{0.999}& \multicolumn{1}{c|}{0.743}& \multicolumn{1}{c}{0.543}&\multicolumn{1}{c}{0.517} & \multicolumn{1}{c}{0.971}& \multicolumn{1}{c}{0.675} \\

    \multicolumn{1}{c}{M19} &\multicolumn{1}{c}{0.497} & \multicolumn{1}{c}{0.198}& \multicolumn{1}{c}{0.400}& \multicolumn{1}{c|}{0.265} & \multicolumn{1}{c}{0.664}&\multicolumn{1}{c}{0.622} & \multicolumn{1}{c}{0.840}& \multicolumn{1}{c|}{0.715}&\multicolumn{1}{c}{0.590}&\multicolumn{1}{c}{0.590} & \multicolumn{1}{c}{0.999}& \multicolumn{1}{c|}{0.742}& \multicolumn{1}{c}{0.554}&\multicolumn{1}{c}{0.524} & \multicolumn{1}{c}{0.973}& \multicolumn{1}{c}{0.681} \\

    \multicolumn{1}{c}{M20} &\multicolumn{1}{c}{0.497} & \multicolumn{1}{c}{0.198}& \multicolumn{1}{c}{0.400}& \multicolumn{1}{c|}{0.265} & \multicolumn{1}{c}{0.692}&\multicolumn{1}{c}{0.639} & \multicolumn{1}{c}{0.884}& \multicolumn{1}{c|}{0.742}&\multicolumn{1}{c}{0.590}&\multicolumn{1}{c}{0.590} & \multicolumn{1}{c}{0.999}& \multicolumn{1}{c|}{0.743}& \multicolumn{1}{c}{0.534}&\multicolumn{1}{c}{0.512} & \multicolumn{1}{c}{0.984}& \multicolumn{1}{c}{0.673} \\

    \multicolumn{1}{c}{M21} &\multicolumn{1}{c}{0.676} & \multicolumn{1}{c}{0.608}& \multicolumn{1}{c}{1.000}& \multicolumn{1}{c|}{0.756} & \multicolumn{1}{c}{0.693}&\multicolumn{1}{c}{0.646} & \multicolumn{1}{c}{0.856}& \multicolumn{1}{c|}{0.736}&\multicolumn{1}{c}{0.590}&\multicolumn{1}{c}{0.590} & \multicolumn{1}{c}{0.999}& \multicolumn{1}{c|}{0.743}& \multicolumn{1}{c}{0.538}&\multicolumn{1}{c}{0.514} & \multicolumn{1}{c}{0.977}& \multicolumn{1}{c}{0.674} \\

    \multicolumn{1}{c}{M22} &\multicolumn{1}{c}{0.497} & \multicolumn{1}{c}{0.198}& \multicolumn{1}{c}{0.400}& \multicolumn{1}{c|}{0.265} & \multicolumn{1}{c}{0.535}&\multicolumn{1}{c}{0.531} & \multicolumn{1}{c}{0.590}& \multicolumn{1}{c|}{0.558}&\multicolumn{1}{c}{0.620}&\multicolumn{1}{c}{0.661} & \multicolumn{1}{c}{0.730}& \multicolumn{1}{c|}{0.694}& \multicolumn{1}{c}{0.746}&\multicolumn{1}{c}{0.701} & \multicolumn{1}{c}{0.839}& \multicolumn{1}{c}{0.763} \\

    \multicolumn{1}{c}{M23} &\multicolumn{1}{c}{0.497} & \multicolumn{1}{c}{0.198}& \multicolumn{1}{c}{0.400}& \multicolumn{1}{c|}{0.265} & \multicolumn{1}{c}{0.580}&\multicolumn{1}{c}{0.569} & \multicolumn{1}{c}{0.669}& \multicolumn{1}{c|}{0.614}&\multicolumn{1}{c}{0.655}&\multicolumn{1}{c}{0.734} & \multicolumn{1}{c}{0.653}& \multicolumn{1}{c|}{0.691}& \multicolumn{1}{c}{0.742}&\multicolumn{1}{c}{0.702} & \multicolumn{1}{c}{0.820}& \multicolumn{1}{c}{0.757} \\

    \multicolumn{1}{c}{M24} &\multicolumn{1}{c}{0.497} & \multicolumn{1}{c}{0.198}& \multicolumn{1}{c}{0.400}& \multicolumn{1}{c|}{0.265} & \multicolumn{1}{c}{0.697}&\multicolumn{1}{c}{0.655} & \multicolumn{1}{c}{0.834}& \multicolumn{1}{c|}{0.734}&\multicolumn{1}{c}{0.637}&\multicolumn{1}{c}{0.649} & \multicolumn{1}{c}{0.842}& \multicolumn{1}{c|}{0.733}& \multicolumn{1}{c}{0.603}&\multicolumn{1}{c}{0.634} & \multicolumn{1}{c}{0.442}& \multicolumn{1}{c}{0.521} \\

    \multicolumn{1}{c}{M25} &\multicolumn{1}{c}{0.497} & \multicolumn{1}{c}{0.198}& \multicolumn{1}{c}{0.400}& \multicolumn{1}{c|}{0.265} & \multicolumn{1}{c}{0.540}&\multicolumn{1}{c}{0.538} & \multicolumn{1}{c}{0.572}& \multicolumn{1}{c|}{0.554}&\multicolumn{1}{c}{0.637}&\multicolumn{1}{c}{0.649} & \multicolumn{1}{c}{0.842}& \multicolumn{1}{c|}{0.733}& \multicolumn{1}{c}{0.745}&\multicolumn{1}{c}{0.705} & \multicolumn{1}{c}{0.822}& \multicolumn{1}{c}{0.759} \\

    \multicolumn{1}{c}{M26} &\multicolumn{1}{c}{0.497} & \multicolumn{1}{c}{0.198}& \multicolumn{1}{c}{0.400}& \multicolumn{1}{c|}{0.265} & \multicolumn{1}{c}{0.686}&\multicolumn{1}{c}{0.663} & \multicolumn{1}{c}{0.757}& \multicolumn{1}{c|}{0.707}&\multicolumn{1}{c}{0.590}&\multicolumn{1}{c}{0.590} & \multicolumn{1}{c}{0.997}& \multicolumn{1}{c|}{0.742}& \multicolumn{1}{c}{0.624}&\multicolumn{1}{c}{0.579} & \multicolumn{1}{c}{0.839}& \multicolumn{1}{c}{0.685} \\

    \multicolumn{1}{c}{M27} &\multicolumn{1}{c}{0.497} & \multicolumn{1}{c}{0.198}& \multicolumn{1}{c}{0.400}& \multicolumn{1}{c|}{0.265} & \multicolumn{1}{c}{0.688}&\multicolumn{1}{c}{0.672} & \multicolumn{1}{c}{0.736}& \multicolumn{1}{c|}{0.702}&\multicolumn{1}{c}{0.589}&\multicolumn{1}{c}{0.590} & \multicolumn{1}{c}{0.997}& \multicolumn{1}{c|}{0.741}& \multicolumn{1}{c}{0.629}&\multicolumn{1}{c}{0.584} & \multicolumn{1}{c}{0.840}& \multicolumn{1}{c}{0.689} \\

    \multicolumn{1}{c}{M28} &\multicolumn{1}{c}{0.497} & \multicolumn{1}{c}{0.198}& \multicolumn{1}{c}{0.400}& \multicolumn{1}{c|}{0.265} & \multicolumn{1}{c}{0.717}&\multicolumn{1}{c}{0.678} & \multicolumn{1}{c}{0.828}& \multicolumn{1}{c|}{0.746}&\multicolumn{1}{c}{0.641}&\multicolumn{1}{c}{0.655} & \multicolumn{1}{c}{0.828}& \multicolumn{1}{c|}{0.731}& \multicolumn{1}{c}{0.617}&\multicolumn{1}{c}{0.668} & \multicolumn{1}{c}{0.428}& \multicolumn{1}{c}{0.521} \\

    \multicolumn{1}{c}{M29} &\multicolumn{1}{c}{0.497} & \multicolumn{1}{c}{0.198}& \multicolumn{1}{c}{0.400}& \multicolumn{1}{c|}{0.265} & \multicolumn{1}{c}{0.714}&\multicolumn{1}{c}{0.679} & \multicolumn{1}{c}{0.813}& \multicolumn{1}{c|}{0.740}&\multicolumn{1}{c}{0.620}&\multicolumn{1}{c}{0.616} & \multicolumn{1}{c}{0.948}& \multicolumn{1}{c|}{0.747}& \multicolumn{1}{c}{0.594}&\multicolumn{1}{c}{0.642} & \multicolumn{1}{c}{0.380}& \multicolumn{1}{c}{0.477} \\

    \multicolumn{1}{c}{M30} &\multicolumn{1}{c}{0.941} & \multicolumn{1}{c}{0.906}& \multicolumn{1}{c}{0.983}& \multicolumn{1}{c|}{0.943} & \multicolumn{1}{c}{0.677}&\multicolumn{1}{c}{0.674} & \multicolumn{1}{c}{0.684}& \multicolumn{1}{c|}{0.679}&\multicolumn{1}{c}{0.590}&\multicolumn{1}{c}{0.590} & \multicolumn{1}{c}{0.999}& \multicolumn{1}{c|}{0.742}& \multicolumn{1}{c}{0.625}&\multicolumn{1}{c}{0.581} & \multicolumn{1}{c}{0.839}& \multicolumn{1}{c}{0.686} \\
    
	\multicolumn{1}{c}{Homogeneous motifs (All)} &\multicolumn{1}{c}{\textbf{0.941}} & \multicolumn{1}{c}{\textbf{0.906}}& \multicolumn{1}{c}{\textbf{0.984}}& \multicolumn{1}{c|}{\textbf{0.943}} & \multicolumn{1}{c}{\textbf{0.783}}&\multicolumn{1}{c}{\textbf{0.743}} & \multicolumn{1}{c}{\textbf{0.867}}& \multicolumn{1}{c|}{\textbf{0.800}}&\multicolumn{1}{c}{\textbf{0.733}}&\multicolumn{1}{c}{\textbf{0.752}} & \multicolumn{1}{c}{\textbf{0.818}}& \multicolumn{1}{c|}{\textbf{0.783}}& \multicolumn{1}{c}{\textbf{0.836}}&\multicolumn{1}{c}{\textbf{0.833}} & \multicolumn{1}{c}{\textbf{0.830}}& \multicolumn{1}{c}{\textbf{0.831}} \\
		\hline
	\end{tabular}
	}
	\label{table:homo}
\end{table*}

\subsubsection{Results on heterogeneous motifs}
To validate the effectiveness of incorporating the heterogeneity of neighborhood preferences, we conduct experiments utilizing heterogeneous motifs. A fusion strategy is introduced, where multiple heterogeneous motifs are combined to train the classifier. Specifically, the likelihood score for each heterogeneous motif, as calculated by Eq. (\ref{equation9}), is treated as a one-dimensional feature. This process generates a 114-dimensional feature set, which is subsequently fed into an XGBoost classifier for bot detection. To provide a comprehensive evaluation, we compare our approach with 10 state-of-the-art methods. Our approach demonstrates a clear advantage over the 10 baseline methods, underscoring its robustness in social bot detection (Table \ref{table:comsxgb}). Specifically, it achieves a 12.4\% improvement in F1 score on Cresci-15 and a 3.60\% increase on TwiBot-20 compared to the best-performing baseline. It also shows strong results across Accuracy, Precision, and Recall, indicating balanced and robust classification capability.

In addition to the above four evaluation metrics, we also use AUC to assess the performance of our method. As shown in Table~\ref{table:aucxgb}, our framework consistently achieves the highest AUC scores across all four datasets, further highlighting its effectiveness. It outperforms all baseline methods, including advanced deep learning models such as BotRGCN and RGT, which rely on multi-modal inputs and transformer-based architectures.

Tables~\ref{table:comsxgb} and \ref{table:aucxgb} collectively demonstrate that our framework achieves the best overall detection performance using only network structure information, highlighting its clear advantages in both resource efficiency and model simplicity. Unlike the strongest baseline, BotRGCN---which relies on a complex relational graph convolutional architecture to fuse network topology, user metadata, and textual content---our approach is significantly more lightweight. It avoids the need to process and store high-dimensional text embeddings or auxiliary user information, making it especially suitable for large-scale or privacy-sensitive scenarios where such data may be unavailable. Furthermore, our framework adopts a theoretically grounded and interpretable feature extraction method based on a Na\"ive Bayes model, followed by standard classifiers. This pipeline is transparent, easy to deploy, and less prone to overfitting compared to deep neural models like BotRGCN, which often require substantial GPU resources and extensive hyperparameter tuning.

Our results demonstrate that heterogeneous motifs significantly outperform homogeneous motifs in detection performance (Table~\ref{table:comsxgb}), reinforcing the importance of incorporating neighborhood heterogeneity to capture diverse interaction patterns within the network. When multiple heterogeneous motifs are combined, the detection performance further improves, indicating that different motif types offer complementary information. Overall, the proposed framework delivers state-of-the-art results while remaining robust and interpretable, offering a compelling alternative to more complex and resource-intensive models.

\begin{table*}[htbp]
    \centering
    \caption{Performance comparison of the Accuracy, Precision, Recall, and F1 across different methods for social bot detection. Here, we report the average performance and standard deviation. For both homogeneous and heterogeneous motifs, results are obtained using an XGBoost classifier. The overall optimal results under the four metrics are marked in bold.} 
    \resizebox{\textwidth}{!}{
    \begin{tabular}{l|rrrrrrrrrrrrrrrr}
    \hline
    \multicolumn{1}{c}{\multirow{2}*{Methods}}& \multicolumn{4}{c|}{Cresci-15}& \multicolumn{4}{c}{MGTAB}\\
     \cline{2-9}
     \multicolumn{1}{c}{}& \multicolumn{1}{c}{Accuracy} & \multicolumn{1}{c}{Precision} & \multicolumn{1}{c}{Recall}& \multicolumn{1}{c|}{F1}& \multicolumn{1}{c}{Accuracy} & \multicolumn{1}{c}{Precision} & \multicolumn{1}{c}{Recall}& \multicolumn{1}{c}{F1}\\
     \hline
     \multicolumn{1}{c}{Ising \cite{des2022detecting}} & \multicolumn{1}{c}{0.554$\pm$0.01} & \multicolumn{1}{c}{0.528$\pm$0.01}& \multicolumn{1}{c}{1.000$\pm$0.00}& \multicolumn{1}{c|}{0.691$\pm$0.02} & \multicolumn{1}{c}{0.539$\pm$0.02}& \multicolumn{1}{c}{0.541$\pm$0.03} & \multicolumn{1}{c}{0.517$\pm$0.02}& \multicolumn{1}{c}{0.528$\pm$0.02}\\
     
     \multicolumn{1}{c}{SybilWalk \cite{jia2017random}} & \multicolumn{1}{c}{0.549$\pm$0.02} & \multicolumn{1}{c}{0.550$\pm$0.02}& \multicolumn{1}{c}{0.998$\pm$0.01}& \multicolumn{1}{c|}{0.710$\pm$0.01} & \multicolumn{1}{c}{0.500$\pm$0.01}& \multicolumn{1}{c}{0.500$\pm$0.03} & \multicolumn{1}{c}{1.000$\pm$0.00}& \multicolumn{1}{c}{0.670$\pm$0.02}\\

     \multicolumn{1}{c}{SybilSCAR \cite{wang2018structure}} & \multicolumn{1}{c}{0.902$\pm$0.01} & \multicolumn{1}{c}{0.990$\pm$0.01}& \multicolumn{1}{c}{0.784$\pm$0.02}& \multicolumn{1}{c|}{0.879$\pm$0.01} & \multicolumn{1}{c}{0.492$\pm$0.01}& \multicolumn{1}{c}{0.167$\pm$0.02} & \multicolumn{1}{c}{0.004$\pm$0.01}& \multicolumn{1}{c}{0.008$\pm$0.02}\\

     \multicolumn{1}{c}{Botometer \cite{yang2022botometer}} &\multicolumn{1}{c}{0.772$\pm$0.01} & \multicolumn{1}{c}{0.809$\pm$0.06}& \multicolumn{1}{c}{0.829$\pm$0.06}& \multicolumn{1}{c|}{0.817$\pm$0.05} & \multicolumn{1}{c}{0.766$\pm$0.01}&\multicolumn{1}{c}{0.762$\pm$0.06} & \multicolumn{1}{c}{0.799$\pm$0.08}& \multicolumn{1}{c}{0.780$\pm$0.05}\\

     \multicolumn{1}{c}{FP \cite{moghaddam2022friendship}} &\multicolumn{1}{c}{0.874$\pm$0.01} & \multicolumn{1}{c}{0.796$\pm$0.07}& \multicolumn{1}{c}{0.832$\pm$0.09}& \multicolumn{1}{c|}{0.809$\pm$0.06} & \multicolumn{1}{c}{0.850$\pm$0.02}&\multicolumn{1}{c}{0.724$\pm$0.06} & \multicolumn{1}{c}{0.706$\pm$0.07}& \multicolumn{1}{c}{0.715$\pm$0.05}\\

     \multicolumn{1}{c}{ARG \cite{abreu2020twitter}} &\multicolumn{1}{c}{0.805$\pm$0.01} & \multicolumn{1}{c}{0.843$\pm$0.06}& \multicolumn{1}{c}{0.829$\pm$0.06}& \multicolumn{1}{c|}{0.835$\pm$0.05} & \multicolumn{1}{c}{0.805$\pm$0.02}&\multicolumn{1}{c}{0.811$\pm$0.09} & \multicolumn{1}{c}{0.781$\pm$0.05}& \multicolumn{1}{c}{0.793$\pm$0.05}\\

     \multicolumn{1}{c}{DeeProBot \cite{hayawi2022deeprobot}} &\multicolumn{1}{c}{0.855$\pm$0.01} & \multicolumn{1}{c}{0.797$\pm$0.07}& \multicolumn{1}{c}{0.788$\pm$0.06}& \multicolumn{1}{c|}{0.790$\pm$0.04} & \multicolumn{1}{c}{0.830$\pm$0.01}&\multicolumn{1}{c}{0.780$\pm$0.08} & \multicolumn{1}{c}{0.845$\pm$0.08}& \multicolumn{1}{c}{0.811$\pm$0.06}\\

     \multicolumn{1}{c}{T5 \cite{raffel2020exploring}} &\multicolumn{1}{c}{0.898$\pm$0.01} & \multicolumn{1}{c}{0.863$\pm$0.06}& \multicolumn{1}{c}{0.845$\pm$0.08}& \multicolumn{1}{c|}{0.850$\pm$0.04} & \multicolumn{1}{c}{0.809$\pm$0.01}&\multicolumn{1}{c}{0.720$\pm$0.07} & \multicolumn{1}{c}{0.734$\pm$0.08}& \multicolumn{1}{c}{0.727$\pm$0.05}\\

     \multicolumn{1}{c}{BotRGCN \cite{feng2021botrgcn}} &\multicolumn{1}{c}{0.973$\pm$0.02} & \multicolumn{1}{c}{0.845$\pm$0.08}& \multicolumn{1}{c}{0.834$\pm$0.07}& \multicolumn{1}{c|}{0.837$\pm$0.05} & \multicolumn{1}{c}{0.838$\pm$0.01}&\multicolumn{1}{c}{0.707$\pm$0.08} & \multicolumn{1}{c}{0.737$\pm$0.09}& \multicolumn{1}{c}{0.721$\pm$0.06}\\

     \multicolumn{1}{c}{RGT \cite{feng2022heterogeneity}} &\multicolumn{1}{c}{0.973$\pm$0.01} & \multicolumn{1}{c}{0.869$\pm$0.07}& \multicolumn{1}{c}{0.840$\pm$0.07}& \multicolumn{1}{c|}{0.851$\pm$0.04} & \multicolumn{1}{c}{0.827$\pm$0.02}&\multicolumn{1}{c}{0.777$\pm$0.05} & \multicolumn{1}{c}{0.702$\pm$0.07}& \multicolumn{1}{c}{0.738$\pm$0.05}\\

     \multicolumn{1}{c}{Homogeneous motifs (All)} &\multicolumn{1}{c}{0.941$\pm$0.01} & \multicolumn{1}{c}{0.906$\pm$0.02}& \multicolumn{1}{c}{0.984$\pm$0.02}& \multicolumn{1}{c|}{0.943$\pm$0.01} & \multicolumn{1}{c}{0.783$\pm$0.03}&\multicolumn{1}{c}{0.743$\pm$0.02} & \multicolumn{1}{c}{0.867$\pm$0.04}& \multicolumn{1}{c}{0.800$\pm$0.02}\\

     \multicolumn{1}{c}{Heterogeneous motifs (All)} &\multicolumn{1}{c}{\textbf{0.987$\pm$0.01}} & \multicolumn{1}{c}{\textbf{0.976$\pm$0.02}}& \multicolumn{1}{c}{\textbf{1.000$\pm$0.00}}& \multicolumn{1}{c|}{\textbf{0.988$\pm$0.01}} & \multicolumn{1}{c}{\textbf{0.826$\pm$0.02}}&\multicolumn{1}{c}{\textbf{0.805$\pm$0.02}} & \multicolumn{1}{c}{\textbf{0.860$\pm$0.02}}& \multicolumn{1}{c}{\textbf{0.831$\pm$0.02}}\\

     \multicolumn{1}{c}{Heterogeneous motifs ($\text{AUC}^{\prime}_\text{upper}>0.7$)} &\multicolumn{1}{c}{\textbf{0.987$\pm$0.01}} & \multicolumn{1}{c}{\textbf{0.976$\pm$0.02}}& \multicolumn{1}{c}{\textbf{1.000$\pm$0.00}}& \multicolumn{1}{c|}{\textbf{0.988$\pm$0.01}} & \multicolumn{1}{c}{0.821$\pm$0.01}&\multicolumn{1}{c}{0.797$\pm$0.02} & \multicolumn{1}{c}{0.863$\pm$0.02}& \multicolumn{1}{c}{0.828$\pm$0.01}\\
     \hline
    \multicolumn{1}{c}{\multirow{2}*{Methods}}& \multicolumn{4}{c|}{TwiBot-20}& \multicolumn{4}{c}{TwiBot-22}\\
    \cline{2-9}
    \multicolumn{1}{c}{}&\multicolumn{1}{c}{Accuracy} & \multicolumn{1}{c}{Precision} & \multicolumn{1}{c}{Recall}& \multicolumn{1}{c|}{F1}&\multicolumn{1}{c}{Accuracy} & \multicolumn{1}{c}{Precision} & \multicolumn{1}{c}{Recall}& \multicolumn{1}{c}{F1}\\
    \hline
    \multicolumn{1}{c}{Ising \cite{des2022detecting}} &\multicolumn{1}{c}{0.573$\pm$0.01}& \multicolumn{1}{c}{0.587$\pm$0.03} & \multicolumn{1}{c}{0.857$\pm$0.02}& \multicolumn{1}{c|}{0.697$\pm$0.01}& \multicolumn{1}{c}{0.493$\pm$0.02}& \multicolumn{1}{c}{0.491$\pm$0.02} & \multicolumn{1}{c}{0.823$\pm$0.02}& \multicolumn{1}{c}{0.615$\pm$0.01} \\

    \multicolumn{1}{c}{SybilWalk \cite{jia2017random}} & \multicolumn{1}{c}{0.678$\pm$0.02}& \multicolumn{1}{c}{0.653$\pm$0.01} & \multicolumn{1}{c}{0.480$\pm$0.02}& \multicolumn{1}{c|}{0.553$\pm$0.03}& \multicolumn{1}{c}{0.537$\pm$0.01}& \multicolumn{1}{c}{0.537$\pm$0.02} & \multicolumn{1}{c}{1.000$\pm$0.00}& \multicolumn{1}{c}{0.699$\pm$0.02} \\

    \multicolumn{1}{c}{SybilSCAR \cite{wang2018structure}} &\multicolumn{1}{c}{0.473$\pm$0.02}& \multicolumn{1}{c}{0.400$\pm$0.03} & \multicolumn{1}{c}{0.466$\pm$0.01}& \multicolumn{1}{c|}{0.430$\pm$0.01}& \multicolumn{1}{c}{0.481$\pm$0.02}& \multicolumn{1}{c}{0.510$\pm$0.02} & \multicolumn{1}{c}{0.897$\pm$0.02}& \multicolumn{1}{c}{0.650$\pm$0.01} \\

    \multicolumn{1}{c}{Botometer \cite{yang2022botometer}} &\multicolumn{1}{c}{0.742$\pm$0.02}&\multicolumn{1}{c}{0.794$\pm$0.07} & \multicolumn{1}{c}{0.773$\pm$0.06}& \multicolumn{1}{c|}{0.783$\pm$0.04}& \multicolumn{1}{c}{0.750$\pm$0.01}&\multicolumn{1}{c}{0.820$\pm$0.06} & \multicolumn{1}{c}{0.837$\pm$0.06}& \multicolumn{1}{c}{0.825$\pm$0.04} \\

    \multicolumn{1}{c}{FP \cite{moghaddam2022friendship}} &\multicolumn{1}{c}{0.817$\pm$0.01}&\multicolumn{1}{c}{0.840$\pm$0.07} & \multicolumn{1}{c}{0.828$\pm$0.05}& \multicolumn{1}{c|}{0.831$\pm$0.03}& \multicolumn{1}{c}{0.811$\pm$0.01}&\multicolumn{1}{c}{0.830$\pm$0.06} & \multicolumn{1}{c}{0.841$\pm$0.08}& \multicolumn{1}{c}{0.832$\pm$0.05} \\

    \multicolumn{1}{c}{ARG \cite{abreu2020twitter}} &\multicolumn{1}{c}{0.803$\pm$0.02}&\multicolumn{1}{c}{0.834$\pm$0.08} & \multicolumn{1}{c}{0.830$\pm$0.08}& \multicolumn{1}{c|}{0.827$\pm$0.05}& \multicolumn{1}{c}{0.801$\pm$0.02}&\multicolumn{1}{c}{0.852$\pm$0.07} & \multicolumn{1}{c}{0.824$\pm$0.06}& \multicolumn{1}{c}{0.835$\pm$0.05} \\

    \multicolumn{1}{c}{DeeProBot \cite{hayawi2022deeprobot}} &\multicolumn{1}{c}{0.826$\pm$0.02}&\multicolumn{1}{c}{0.771$\pm$0.07} & \multicolumn{1}{c}{0.733$\pm$0.07}& \multicolumn{1}{c|}{0.751$\pm$0.07}& \multicolumn{1}{c}{0.833$\pm$0.02}&\multicolumn{1}{c}{0.783$\pm$0.06} & \multicolumn{1}{c}{0.841$\pm$0.08}& \multicolumn{1}{c}{0.807$\pm$0.04} \\

    \multicolumn{1}{c}{T5 \cite{raffel2020exploring}} &\multicolumn{1}{c}{0.819$\pm$0.02}&\multicolumn{1}{c}{0.729$\pm$0.07} & \multicolumn{1}{c}{0.867$\pm$0.06}& \multicolumn{1}{c|}{0.792$\pm$0.05}& \multicolumn{1}{c}{0.846$\pm$0.01}&\multicolumn{1}{c}{0.840$\pm$0.08} & \multicolumn{1}{c}{0.820$\pm$0.05}& \multicolumn{1}{c}{0.828$\pm$0.06} \\

    \multicolumn{1}{c}{BotRGCN \cite{feng2021botrgcn}} &\multicolumn{1}{c}{0.817$\pm$0.01}&\multicolumn{1}{c}{0.827$\pm$0.06} & \multicolumn{1}{c}{0.840$\pm$0.07}& \multicolumn{1}{c|}{0.833$\pm$0.05}& \multicolumn{1}{c}{0.851$\pm$0.01}&\multicolumn{1}{c}{0.715$\pm$0.08} & \multicolumn{1}{c}{0.808$\pm$0.07}& \multicolumn{1}{c}{0.759$\pm$0.05} \\

    \multicolumn{1}{c}{RGT \cite{feng2022heterogeneity}} &\multicolumn{1}{c}{0.844$\pm$0.01}&\multicolumn{1}{c}{0.720$\pm$0.09} & \multicolumn{1}{c}{0.892$\pm$0.07}& \multicolumn{1}{c|}{0.797$\pm$0.05}& \multicolumn{1}{c}{0.852$\pm$0.02}&\multicolumn{1}{c}{0.845$\pm$0.06} & \multicolumn{1}{c}{0.815$\pm$0.08}& \multicolumn{1}{c}{0.827$\pm$0.05} \\

    \multicolumn{1}{c}{Homogeneous motifs (All)} &\multicolumn{1}{c}{0.733$\pm$0.01}&\multicolumn{1}{c}{0.752$\pm$0.01} & \multicolumn{1}{c}{0.818$\pm$0.02}& \multicolumn{1}{c|}{0.783$\pm$0.01}& \multicolumn{1}{c}{0.836$\pm$0.01}&\multicolumn{1}{c}{0.833$\pm$0.01} & \multicolumn{1}{c}{0.830$\pm$0.01}& \multicolumn{1}{c}{0.831$\pm$0.01} \\

    \multicolumn{1}{c}{Heterogeneous motifs (All)} &\multicolumn{1}{c}{\textbf{0.827$\pm$0.01}}&\multicolumn{1}{c}{\textbf{0.812$\pm$0.01}} & \multicolumn{1}{c}{\textbf{0.921$\pm$0.01}}& \multicolumn{1}{c|}{\textbf{0.863$\pm$0.01}}& \multicolumn{1}{c}{\textbf{0.874$\pm$0.00}}&\multicolumn{1}{c}{\textbf{0.870$\pm$0.00}} & \multicolumn{1}{c}{\textbf{0.873$\pm$0.00}}& \multicolumn{1}{c}{\textbf{0.872$\pm$0.00}} \\

    \multicolumn{1}{c}{Heterogeneous motifs ($\text{AUC}^{\prime}_\text{upper}>0.7$)} &\multicolumn{1}{c}{0.749$\pm$0.01}&\multicolumn{1}{c}{0.742$\pm$0.01} & \multicolumn{1}{c}{0.882$\pm$0.02}& \multicolumn{1}{c|}{0.806$\pm$0.01}& \multicolumn{1}{c}{0.857$\pm$0.00}&\multicolumn{1}{c}{0.852$\pm$0.00} & \multicolumn{1}{c}{0.856$\pm$0.00}& \multicolumn{1}{c}{0.854$\pm$0.00} \\
    \hline
    \end{tabular}
    }
    \label{table:comsxgb}
\end{table*}

\begin{table*}[htbp]
    \centering
\caption{Performance comparison of the AUC across different methods for social bot detection. Here, we report the average performance and standard deviation. For both homogeneous and heterogeneous motifs, results are obtained using an XGBoost classifier. The optimal results are marked in bold.}
\begin{tabular}{cccccc}
\hline
Datasets & Cresci-15 & MGTAB & TwiBot-20 & TwiBot-22 \\
\hline
Ising \cite{des2022detecting} &0.553$\pm$0.02& 0.539$\pm$0.02 & 0.653$\pm$0.03 & 0.620$\pm$0.01 \\
SybilWalk \cite{jia2017random}  & 0.500$\pm$0.01 & 0.500$\pm$0.01 & 0.558$\pm$0.02 & 0.673$\pm$0.02 \\
SybilSCAR \cite{wang2018structure}  & 0.892$\pm$0.03 & 0.492$\pm$0.02 & 0.420$\pm$0.01 & 0.639$\pm$0.01\\
Botometer \cite{yang2022botometer} & 0.823$\pm$0.01 & 0.810$\pm$0.01 & 0.810$\pm$0.02 & 0.810$\pm$0.01 \\
FP \cite{moghaddam2022friendship} & 0.848$\pm$0.01 & 0.817$\pm$0.02 & 0.797$\pm$0.01 & 0.792$\pm$0.01 \\
ARG \cite{abreu2020twitter} & 0.838$\pm0.03$ & 0.807$\pm$0.01 & 0.809$\pm0.03$ & 0.813$\pm$0.02 \\
DeeProBot \cite{hayawi2022deeprobot} & 0.885$\pm0.01$ & 0.855$\pm$0.01 & 0.867$\pm$0.02 & 0.863$\pm$0.02 \\
T5 \cite{raffel2020exploring}& 0.855$\pm$0.01 & 0.835$\pm$0.01 & 0.841$\pm$0.02 & 0.857$\pm$0.01 \\
BotRGCN \cite{feng2021botrgcn}& 0.951$\pm$0.02 & 0.851$\pm$0.01 & 0.866$\pm$0.01 & 0.896$\pm$0.01 \\
RGT \cite{feng2022heterogeneity}& 0.915$\pm$0.01 & 0.842$\pm$0.02 & 0.848$\pm$0.02 & 0.862$\pm$0.02 \\
Homogeneous motifs (All)& 0.972$\pm$0.01 & 0.851$\pm$0.02 & 0.787$\pm$0.01 & 0.908$\pm$0.00 \\
Heterogeneous motifs (All)& \textbf{0.992$\pm$0.01} & \textbf{0.907$\pm$0.02} & \textbf{0.914$\pm$0.01} & \textbf{0.942$\pm$0.00} \\
Heterogeneous motifs ($\text{AUC}^{\prime}_\text{upper}>0.7$)& 0.987$\pm$0.01 & 0.904$\pm$0.01 & 0.846$\pm$0.01 & 0.929$\pm$0.00 \\
\hline
\end{tabular}
\label{table:aucxgb}
\end{table*}

\subsubsection{Robustness analysis of other machine learning algorithms}
The homogeneous and heterogeneous motifs proposed in this study, derived from a Na\"ive Bayes model, are broadly applicable and compatible with a wide range of machine learning classifiers. To further validate the robustness of these features, we also evaluate their performance using Random Forest \cite{yang2020scalable, shi2025mgtab} and Gradient Boosting \cite{mohanty2022enhancing}. As shown in Tables \ref{table:coms}-\ref{table:aucgb}, the detection performance remains stable across the other classifiers, indicating that the proposed motif-based features are not sensitive to the choice of algorithm. Our approach consistently outperforms state-of-the-art methods across five standard evaluation metrics, further demonstrating the generalizability and effectiveness of the proposed framework.

Although the proposed heterogeneous motifs are broadly applicable and consistently perform well across various machine learning classifiers, our experiments with XGBoost, Random Forest, and Gradient Boosting reveal that detection performance still exhibits slight variations across models (Tables \ref{table:comsxgb}-\ref{table:aucgb}). These findings suggest that carefully selecting an efficient classifier can, in some cases, further enhance detection performance. In this study, the proposed heterogeneous motifs yield the best detection performance when implemented with the XGBoost classifier (Tables \ref{table:comsxgb} and \ref{table:aucxgb}).
 
\begin{table*}[htbp]
    \centering
    \caption{Performance comparison of the Accuracy, Precision, Recall, and F1 across different methods for social bot detection. Here, we report the average performance and standard deviation. For both homogeneous and heterogeneous motifs, results are obtained using a Random Forest classifier. The overall optimal results under the four metrics are marked in bold.}
    \resizebox{\textwidth}{!}{
    \begin{tabular}{l|rrrrrrrrrrrrrrrr}
    \hline
    \multicolumn{1}{c}{\multirow{2}*{Methods}}& \multicolumn{4}{c|}{Cresci-15}& \multicolumn{4}{c}{MGTAB}\\
     \cline{2-9}
     \multicolumn{1}{c}{}& \multicolumn{1}{c}{Accuracy} & \multicolumn{1}{c}{Precision} & \multicolumn{1}{c}{Recall}& \multicolumn{1}{c|}{F1}& \multicolumn{1}{c}{Accuracy} & \multicolumn{1}{c}{Precision} & \multicolumn{1}{c}{Recall}& \multicolumn{1}{c}{F1}\\
     \hline
     \multicolumn{1}{c}{Ising \cite{des2022detecting}} & \multicolumn{1}{c}{0.554$\pm$0.01} & \multicolumn{1}{c}{0.528$\pm$0.01}& \multicolumn{1}{c}{1.000$\pm$0.00}& \multicolumn{1}{c|}{0.691$\pm$0.02} & \multicolumn{1}{c}{0.539$\pm$0.02}& \multicolumn{1}{c}{0.541$\pm$0.03} & \multicolumn{1}{c}{0.517$\pm$0.02}& \multicolumn{1}{c}{0.528$\pm$0.02}\\
     
     \multicolumn{1}{c}{SybilWalk \cite{jia2017random}} & \multicolumn{1}{c}{0.549$\pm$0.02} & \multicolumn{1}{c}{0.550$\pm$0.02}& \multicolumn{1}{c}{0.998$\pm$0.01}& \multicolumn{1}{c|}{0.710$\pm$0.01} & \multicolumn{1}{c}{0.500$\pm$0.01}& \multicolumn{1}{c}{0.500$\pm$0.03} & \multicolumn{1}{c}{1.000$\pm$0.00}& \multicolumn{1}{c}{0.670$\pm$0.02}\\

     \multicolumn{1}{c}{SybilSCAR \cite{wang2018structure}} & \multicolumn{1}{c}{0.902$\pm$0.01} & \multicolumn{1}{c}{0.990$\pm$0.01}& \multicolumn{1}{c}{0.784$\pm$0.02}& \multicolumn{1}{c|}{0.879$\pm$0.01} & \multicolumn{1}{c}{0.492$\pm$0.01}& \multicolumn{1}{c}{0.167$\pm$0.02} & \multicolumn{1}{c}{0.004$\pm$0.01}& \multicolumn{1}{c}{0.008$\pm$0.02}\\

     \multicolumn{1}{c}{Botometer \cite{yang2022botometer}} &\multicolumn{1}{c}{0.772$\pm$0.01} & \multicolumn{1}{c}{0.809$\pm$0.06}& \multicolumn{1}{c}{0.829$\pm$0.06}& \multicolumn{1}{c|}{0.817$\pm$0.05} & \multicolumn{1}{c}{0.766$\pm$0.01}&\multicolumn{1}{c}{0.762$\pm$0.06} & \multicolumn{1}{c}{0.799$\pm$0.08}& \multicolumn{1}{c}{0.780$\pm$0.05}\\

     \multicolumn{1}{c}{FP \cite{moghaddam2022friendship}} &\multicolumn{1}{c}{0.874$\pm$0.01} & \multicolumn{1}{c}{0.796$\pm$0.07}& \multicolumn{1}{c}{0.832$\pm$0.09}& \multicolumn{1}{c|}{0.809$\pm$0.06} & \multicolumn{1}{c}{0.850$\pm$0.02}&\multicolumn{1}{c}{0.724$\pm$0.06} & \multicolumn{1}{c}{0.706$\pm$0.07}& \multicolumn{1}{c}{0.715$\pm$0.05}\\

     \multicolumn{1}{c}{ARG \cite{abreu2020twitter}} &\multicolumn{1}{c}{0.805$\pm$0.01} & \multicolumn{1}{c}{0.843$\pm$0.06}& \multicolumn{1}{c}{0.829$\pm$0.06}& \multicolumn{1}{c|}{0.835$\pm$0.05} & \multicolumn{1}{c}{0.805$\pm$0.02}&\multicolumn{1}{c}{0.811$\pm$0.09} & \multicolumn{1}{c}{0.781$\pm$0.05}& \multicolumn{1}{c}{0.793$\pm$0.05}\\

     \multicolumn{1}{c}{DeeProBot \cite{hayawi2022deeprobot}} &\multicolumn{1}{c}{0.855$\pm$0.01} & \multicolumn{1}{c}{0.797$\pm$0.07}& \multicolumn{1}{c}{0.788$\pm$0.06}& \multicolumn{1}{c|}{0.790$\pm$0.04} & \multicolumn{1}{c}{0.830$\pm$0.01}&\multicolumn{1}{c}{0.780$\pm$0.08} & \multicolumn{1}{c}{0.845$\pm$0.08}& \multicolumn{1}{c}{0.811$\pm$0.06}\\

     \multicolumn{1}{c}{T5 \cite{raffel2020exploring}} &\multicolumn{1}{c}{0.898$\pm$0.01} & \multicolumn{1}{c}{0.863$\pm$0.06}& \multicolumn{1}{c}{0.845$\pm$0.08}& \multicolumn{1}{c|}{0.850$\pm$0.04} & \multicolumn{1}{c}{0.809$\pm$0.01}&\multicolumn{1}{c}{0.720$\pm$0.07} & \multicolumn{1}{c}{0.734$\pm$0.08}& \multicolumn{1}{c}{0.727$\pm$0.05}\\

     \multicolumn{1}{c}{BotRGCN \cite{feng2021botrgcn}} &\multicolumn{1}{c}{0.973$\pm$0.02} & \multicolumn{1}{c}{0.845$\pm$0.08}& \multicolumn{1}{c}{0.834$\pm$0.07}& \multicolumn{1}{c|}{0.837$\pm$0.05} & \multicolumn{1}{c}{0.838$\pm$0.01}&\multicolumn{1}{c}{0.707$\pm$0.08} & \multicolumn{1}{c}{0.737$\pm$0.09}& \multicolumn{1}{c}{0.721$\pm$0.06}\\

     \multicolumn{1}{c}{RGT \cite{feng2022heterogeneity}} &\multicolumn{1}{c}{0.973$\pm$0.01} & \multicolumn{1}{c}{0.869$\pm$0.07}& \multicolumn{1}{c}{0.840$\pm$0.07}& \multicolumn{1}{c|}{0.851$\pm$0.04} & \multicolumn{1}{c}{0.827$\pm$0.02}&\multicolumn{1}{c}{0.777$\pm$0.05} & \multicolumn{1}{c}{0.702$\pm$0.07}& \multicolumn{1}{c}{0.738$\pm$0.05}\\

     \multicolumn{1}{c}{Homogeneous motifs (All)} &\multicolumn{1}{c}{0.940$\pm$0.02} & \multicolumn{1}{c}{0.905$\pm$0.01}& \multicolumn{1}{c}{0.984$\pm$0.02}& \multicolumn{1}{c|}{0.943$\pm$0.02} & \multicolumn{1}{c}{0.784$\pm$0.01}&\multicolumn{1}{c}{0.737$\pm$0.02} & \multicolumn{1}{c}{0.887$\pm$0.03}& \multicolumn{1}{c}{0.805$\pm$0.02}\\

     \multicolumn{1}{c}{Heterogeneous motifs (All)} &\multicolumn{1}{c}{0.984$\pm$0.01} & \multicolumn{1}{c}{0.970$\pm$0.01}& \multicolumn{1}{c}{0.998$\pm$0.02}& \multicolumn{1}{c|}{0.984$\pm$0.02} & \multicolumn{1}{c}{\textbf{0.807$\pm$0.02}}&\multicolumn{1}{c}{\textbf{0.780$\pm$0.02}} & \multicolumn{1}{c}{\textbf{0.857$\pm$0.01}}& \multicolumn{1}{c}{\textbf{0.816$\pm$0.02}}\\

     \multicolumn{1}{c}{Heterogeneous motifs ($\text{AUC}^{\prime}_\text{upper}>0.7$)} &\multicolumn{1}{c}{\textbf{0.987$\pm$0.01}} & \multicolumn{1}{c}{\textbf{0.976$\pm$0.01}}& \multicolumn{1}{c}{\textbf{1.000$\pm$0.00}}& \multicolumn{1}{c|}{\textbf{0.988$\pm$0.02}} & \multicolumn{1}{c}{0.803$\pm$0.01}&\multicolumn{1}{c}{0.775$\pm$0.02} & \multicolumn{1}{c}{0.853$\pm$0.02}& \multicolumn{1}{c}{0.812$\pm$0.01}\\
     \hline
    \multicolumn{1}{c}{\multirow{2}*{Methods}}& \multicolumn{4}{c|}{TwiBot-20}& \multicolumn{4}{c}{TwiBot-22}\\
    \cline{2-9}
    \multicolumn{1}{c}{}&\multicolumn{1}{c}{Accuracy} & \multicolumn{1}{c}{Precision} & \multicolumn{1}{c}{Recall}& \multicolumn{1}{c|}{F1}&\multicolumn{1}{c}{Accuracy} & \multicolumn{1}{c}{Precision} & \multicolumn{1}{c}{Recall}& \multicolumn{1}{c}{F1}\\
    \hline
    \multicolumn{1}{c}{Ising \cite{des2022detecting}} &\multicolumn{1}{c}{0.573$\pm$0.01}& \multicolumn{1}{c}{0.587$\pm$0.03} & \multicolumn{1}{c}{0.857$\pm$0.02}& \multicolumn{1}{c|}{0.697$\pm$0.01}& \multicolumn{1}{c}{0.493$\pm$0.02}& \multicolumn{1}{c}{0.491$\pm$0.02} & \multicolumn{1}{c}{0.823$\pm$0.02}& \multicolumn{1}{c}{0.615$\pm$0.01} \\

    \multicolumn{1}{c}{SybilWalk \cite{jia2017random}} &\multicolumn{1}{c}{0.678$\pm$0.02}& \multicolumn{1}{c}{0.653$\pm$0.01} & \multicolumn{1}{c}{0.480$\pm$0.02}& \multicolumn{1}{c|}{0.553$\pm$0.03}& \multicolumn{1}{c}{0.537$\pm$0.01}& \multicolumn{1}{c}{0.537$\pm$0.02} & \multicolumn{1}{c}{1.000$\pm$0.00}& \multicolumn{1}{c}{0.699$\pm$0.02} \\

    \multicolumn{1}{c}{SybilSCAR \cite{wang2018structure}} &\multicolumn{1}{c}{0.473$\pm$0.02}& \multicolumn{1}{c}{0.400$\pm$0.03} & \multicolumn{1}{c}{0.466$\pm$0.01}& \multicolumn{1}{c|}{0.430$\pm$0.01}& \multicolumn{1}{c}{0.481$\pm$0.02}& \multicolumn{1}{c}{0.510$\pm$0.02} & \multicolumn{1}{c}{0.897$\pm$0.02}& \multicolumn{1}{c}{0.650$\pm$0.01} \\

    \multicolumn{1}{c}{Botometer \cite{yang2022botometer}} &\multicolumn{1}{c}{0.742$\pm$0.02}&\multicolumn{1}{c}{0.794$\pm$0.07} & \multicolumn{1}{c}{0.773$\pm$0.06}& \multicolumn{1}{c|}{0.783$\pm$0.04}& \multicolumn{1}{c}{0.750$\pm$0.01}&\multicolumn{1}{c}{0.820$\pm$0.06} & \multicolumn{1}{c}{0.837$\pm$0.06}& \multicolumn{1}{c}{0.825$\pm$0.04} \\

    \multicolumn{1}{c}{FP \cite{moghaddam2022friendship}} &\multicolumn{1}{c}{0.817$\pm$0.01}&\multicolumn{1}{c}{0.840$\pm$0.07} & \multicolumn{1}{c}{0.828$\pm$0.05}& \multicolumn{1}{c|}{0.831$\pm$0.03}& \multicolumn{1}{c}{0.811$\pm$0.01}&\multicolumn{1}{c}{0.830$\pm$0.06} & \multicolumn{1}{c}{0.841$\pm$0.08}& \multicolumn{1}{c}{0.832$\pm$0.05} \\

    \multicolumn{1}{c}{ARG \cite{abreu2020twitter}} &\multicolumn{1}{c}{0.803$\pm$0.02}&\multicolumn{1}{c}{0.834$\pm$0.08} & \multicolumn{1}{c}{0.830$\pm$0.08}& \multicolumn{1}{c|}{0.827$\pm$0.05}& \multicolumn{1}{c}{0.801$\pm$0.02}&\multicolumn{1}{c}{0.852$\pm$0.07} & \multicolumn{1}{c}{0.824$\pm$0.06}& \multicolumn{1}{c}{0.835$\pm$0.05} \\

    \multicolumn{1}{c}{DeeProBot \cite{hayawi2022deeprobot}} &\multicolumn{1}{c}{0.826$\pm$0.02}&\multicolumn{1}{c}{0.771$\pm$0.07} & \multicolumn{1}{c}{0.733$\pm$0.07}& \multicolumn{1}{c|}{0.751$\pm$0.07}& \multicolumn{1}{c}{0.833$\pm$0.02}&\multicolumn{1}{c}{0.783$\pm$0.06} & \multicolumn{1}{c}{0.841$\pm$0.08}& \multicolumn{1}{c}{0.807$\pm$0.04} \\

    \multicolumn{1}{c}{T5 \cite{raffel2020exploring}} &\multicolumn{1}{c}{0.819$\pm$0.02}&\multicolumn{1}{c}{0.729$\pm$0.07} & \multicolumn{1}{c}{0.867$\pm$0.06}& \multicolumn{1}{c|}{0.792$\pm$0.05}& \multicolumn{1}{c}{0.846$\pm$0.01}&\multicolumn{1}{c}{0.840$\pm$0.08} & \multicolumn{1}{c}{0.820$\pm$0.05}& \multicolumn{1}{c}{0.828$\pm$0.06} \\

    \multicolumn{1}{c}{BotRGCN \cite{feng2021botrgcn}} &\multicolumn{1}{c}{0.817$\pm$0.01}&\multicolumn{1}{c}{0.827$\pm$0.06} & \multicolumn{1}{c}{0.840$\pm$0.07}& \multicolumn{1}{c|}{0.833$\pm$0.05}& \multicolumn{1}{c}{0.851$\pm$0.01}&\multicolumn{1}{c}{0.715$\pm$0.08} & \multicolumn{1}{c}{0.808$\pm$0.07}& \multicolumn{1}{c}{0.759$\pm$0.05} \\

    \multicolumn{1}{c}{RGT \cite{feng2022heterogeneity}} &\multicolumn{1}{c}{0.844$\pm$0.01}&\multicolumn{1}{c}{0.720$\pm$0.09} & \multicolumn{1}{c}{0.892$\pm$0.07}& \multicolumn{1}{c|}{0.797$\pm$0.05}& \multicolumn{1}{c}{0.852$\pm$0.02}&\multicolumn{1}{c}{0.845$\pm$0.06} & \multicolumn{1}{c}{0.815$\pm$0.08}& \multicolumn{1}{c}{0.827$\pm$0.05} \\

    \multicolumn{1}{c}{Homogeneous motifs (All)} &\multicolumn{1}{c}{0.730$\pm$0.02}&\multicolumn{1}{c}{0.744$\pm$0.01} & \multicolumn{1}{c}{0.827$\pm$0.02}& \multicolumn{1}{c|}{0.783$\pm$0.02}& \multicolumn{1}{c}{0.834$\pm$0.02}&\multicolumn{1}{c}{0.834$\pm$0.01} & \multicolumn{1}{c}{0.825$\pm$0.02}& \multicolumn{1}{c}{0.829$\pm$0.02} \\

    \multicolumn{1}{c}{Heterogeneous motifs (All)} &\multicolumn{1}{c}{\textbf{0.825$\pm$0.01}}&\multicolumn{1}{c}{\textbf{0.809$\pm$0.02}} & \multicolumn{1}{c}{\textbf{0.921$\pm$0.02}}& \multicolumn{1}{c|}{\textbf{0.862$\pm$0.01}}& \multicolumn{1}{c}{0.839$\pm$0.01}&\multicolumn{1}{c}{0.834$\pm$0.02} & \multicolumn{1}{c}{0.836$\pm$0.02}& \multicolumn{1}{c}{0.835$\pm$0.01} \\

    \multicolumn{1}{c}{Heterogeneous motifs ($\text{AUC}^{\prime}_\text{upper}>0.7$)} &\multicolumn{1}{c}{0.750$\pm$0.01}&\multicolumn{1}{c}{0.743$\pm$0.02} & \multicolumn{1}{c}{0.882$\pm$0.02}& \multicolumn{1}{c|}{0.807$\pm$0.01}& \multicolumn{1}{c}{\textbf{0.845$\pm$0.01}}&\multicolumn{1}{c}{\textbf{0.846$\pm$0.02}} & \multicolumn{1}{c}{\textbf{0.836$\pm$0.01}}& \multicolumn{1}{c}{\textbf{0.841$\pm$0.02}} \\
    \hline
    \end{tabular}
    }
    \label{table:coms}
\end{table*}

\begin{table*}[htbp]
\centering
\caption{Performance comparison of the AUC across different methods for social bot detection. Here, we report the average performance and standard deviation. For both homogeneous and heterogeneous motifs, results are obtained using a Random Forest classifier. The optimal results are marked in bold.}
\begin{tabular}{cccccc}
\hline
Datasets & Cresci-15 & MGTAB & TwiBot-20 & TwiBot-22 \\
\hline
Ising \cite{des2022detecting} &0.553$\pm$0.02& 0.539$\pm$0.02 & 0.653$\pm$0.03 & 0.620$\pm$0.01 \\
SybilWalk \cite{jia2017random}  & 0.500$\pm$0.01 & 0.500$\pm$0.01 & 0.558$\pm$0.02 & 0.673$\pm$0.02 \\
SybilSCAR \cite{wang2018structure}  & 0.892$\pm$0.03 & 0.492$\pm$0.02 & 0.420$\pm$0.01 & 0.639$\pm$0.01\\
Botometer \cite{yang2022botometer} & 0.823$\pm$0.01 & 0.810$\pm$0.01 & 0.810$\pm$0.02 & 0.810$\pm$0.01 \\
FP \cite{moghaddam2022friendship} & 0.848$\pm$0.01 & 0.817$\pm$0.02 & 0.797$\pm$0.01 & 0.792$\pm$0.01 \\
ARG \cite{abreu2020twitter} & 0.838$\pm0.03$ & 0.807$\pm$0.01 & 0.809$\pm0.03$ & 0.813$\pm$0.02 \\
DeeProBot \cite{hayawi2022deeprobot} & 0.885$\pm0.01$ & 0.855$\pm$0.01 & 0.867$\pm$0.02 & 0.863$\pm$0.02 \\
T5 \cite{raffel2020exploring}& 0.855$\pm$0.01 & 0.835$\pm$0.01 & 0.841$\pm$0.02 & 0.857$\pm$0.01 \\
BotRGCN \cite{feng2021botrgcn}& 0.951$\pm$0.02 & 0.851$\pm$0.01 & 0.866$\pm$0.01 & 0.896$\pm$0.01 \\
RGT \cite{feng2022heterogeneity}& 0.915$\pm$0.01 & 0.842$\pm$0.02 & 0.848$\pm$0.02 & 0.862$\pm$0.02 \\
Homogeneous motifs (All)& 0.975$\pm$0.01 & 0.847$\pm$0.02 & 0.787$\pm$0.01 & 0.919$\pm$0.02 \\
Heterogeneous motifs (All)& \textbf{0.992$\pm$0.02} & \textbf{0.890$\pm$0.01} & \textbf{0.905$\pm$0.02} & 0.922$\pm$0.01 \\
Heterogeneous motifs ($\text{AUC}^{\prime}_\text{upper}>0.7$)& 0.987$\pm$0.02 & \textbf{0.890$\pm$0.02} & 0.848$\pm$0.01 & \textbf{0.929$\pm$0.02} \\
\hline
\end{tabular}
\label{table:auc}
\end{table*}

\begin{table*}[htbp]
    \centering
    \caption{Performance comparison of the Accuracy, Precision, Recall, and F1 across different methods for social bot detection. Here, we report the average performance and standard deviation. For both homogeneous and heterogeneous motifs, results are obtained using a Gradient Boosting classifier. The overall optimal results under the four metrics are marked in bold.} 
    \resizebox{\textwidth}{!}{
    \begin{tabular}{l|rrrrrrrrrrrrrrrr}
    \hline
    \multicolumn{1}{c}{\multirow{2}*{Methods}}& \multicolumn{4}{c|}{Cresci-15}& \multicolumn{4}{c}{MGTAB}\\
     \cline{2-9}
     \multicolumn{1}{c}{}& \multicolumn{1}{c}{Accuracy} & \multicolumn{1}{c}{Precision} & \multicolumn{1}{c}{Recall}& \multicolumn{1}{c|}{F1}& \multicolumn{1}{c}{Accuracy} & \multicolumn{1}{c}{Precision} & \multicolumn{1}{c}{Recall}& \multicolumn{1}{c}{F1}\\
     \hline
     \multicolumn{1}{c}{Ising \cite{des2022detecting}} & \multicolumn{1}{c}{0.554$\pm$0.01} & \multicolumn{1}{c}{0.528$\pm$0.01}& \multicolumn{1}{c}{1.000$\pm$0.00}& \multicolumn{1}{c|}{0.691$\pm$0.02} & \multicolumn{1}{c}{0.539$\pm$0.02}& \multicolumn{1}{c}{0.541$\pm$0.03} & \multicolumn{1}{c}{0.517$\pm$0.02}& \multicolumn{1}{c}{0.528$\pm$0.02}\\

     \multicolumn{1}{c}{SybilWalk \cite{jia2017random}} & \multicolumn{1}{c}{0.549$\pm$0.02} & \multicolumn{1}{c}{0.550$\pm$0.02}& \multicolumn{1}{c}{0.998$\pm$0.01}& \multicolumn{1}{c|}{0.710$\pm$0.01} & \multicolumn{1}{c}{0.500$\pm$0.01}& \multicolumn{1}{c}{0.500$\pm$0.03} & \multicolumn{1}{c}{1.000$\pm$0.00}& \multicolumn{1}{c}{0.670$\pm$0.02}\\

     \multicolumn{1}{c}{SybilSCAR \cite{wang2018structure}} & \multicolumn{1}{c}{0.902$\pm$0.01} & \multicolumn{1}{c}{0.990$\pm$0.01}& \multicolumn{1}{c}{0.784$\pm$0.02}& \multicolumn{1}{c|}{0.879$\pm$0.01} & \multicolumn{1}{c}{0.492$\pm$0.01}& \multicolumn{1}{c}{0.167$\pm$0.02} & \multicolumn{1}{c}{0.004$\pm$0.01}& \multicolumn{1}{c}{0.008$\pm$0.02}\\

     \multicolumn{1}{c}{Botometer \cite{yang2022botometer}} &\multicolumn{1}{c}{0.772$\pm$0.01} & \multicolumn{1}{c}{0.809$\pm$0.06}& \multicolumn{1}{c}{0.829$\pm$0.06}& \multicolumn{1}{c|}{0.817$\pm$0.05} & \multicolumn{1}{c}{0.766$\pm$0.01}&\multicolumn{1}{c}{0.762$\pm$0.06} & \multicolumn{1}{c}{0.799$\pm$0.08}& \multicolumn{1}{c}{0.780$\pm$0.05}\\

     \multicolumn{1}{c}{FP \cite{moghaddam2022friendship}} &\multicolumn{1}{c}{0.874$\pm$0.01} & \multicolumn{1}{c}{0.796$\pm$0.07}& \multicolumn{1}{c}{0.832$\pm$0.09}& \multicolumn{1}{c|}{0.809$\pm$0.06} & \multicolumn{1}{c}{0.850$\pm$0.02}&\multicolumn{1}{c}{0.724$\pm$0.06} & \multicolumn{1}{c}{0.706$\pm$0.07}& \multicolumn{1}{c}{0.715$\pm$0.05}\\

     \multicolumn{1}{c}{ARG \cite{abreu2020twitter}} &\multicolumn{1}{c}{0.805$\pm$0.01} & \multicolumn{1}{c}{0.843$\pm$0.06}& \multicolumn{1}{c}{0.829$\pm$0.06}& \multicolumn{1}{c|}{0.835$\pm$0.05} & \multicolumn{1}{c}{0.805$\pm$0.02}&\multicolumn{1}{c}{0.811$\pm$0.09} & \multicolumn{1}{c}{0.781$\pm$0.05}& \multicolumn{1}{c}{0.793$\pm$0.05}\\

     \multicolumn{1}{c}{DeeProBot \cite{hayawi2022deeprobot}} &\multicolumn{1}{c}{0.855$\pm$0.01} & \multicolumn{1}{c}{0.797$\pm$0.07}& \multicolumn{1}{c}{0.788$\pm$0.06}& \multicolumn{1}{c|}{0.790$\pm$0.04} & \multicolumn{1}{c}{0.830$\pm$0.01}&\multicolumn{1}{c}{0.780$\pm$0.08} & \multicolumn{1}{c}{0.845$\pm$0.08}& \multicolumn{1}{c}{0.811$\pm$0.06}\\

     \multicolumn{1}{c}{T5 \cite{raffel2020exploring}} &\multicolumn{1}{c}{0.898$\pm$0.01} & \multicolumn{1}{c}{0.863$\pm$0.06}& \multicolumn{1}{c}{0.845$\pm$0.08}& \multicolumn{1}{c|}{0.850$\pm$0.04} & \multicolumn{1}{c}{0.809$\pm$0.01}&\multicolumn{1}{c}{0.720$\pm$0.07} & \multicolumn{1}{c}{0.734$\pm$0.08}& \multicolumn{1}{c}{0.727$\pm$0.05}\\

     \multicolumn{1}{c}{BotRGCN \cite{feng2021botrgcn}} &\multicolumn{1}{c}{0.973$\pm$0.02} & \multicolumn{1}{c}{0.845$\pm$0.08}& \multicolumn{1}{c}{0.834$\pm$0.07}& \multicolumn{1}{c|}{0.837$\pm$0.05} & \multicolumn{1}{c}{0.838$\pm$0.01}&\multicolumn{1}{c}{0.707$\pm$0.08} & \multicolumn{1}{c}{0.737$\pm$0.09}& \multicolumn{1}{c}{0.721$\pm$0.06}\\

     \multicolumn{1}{c}{RGT \cite{feng2022heterogeneity}} &\multicolumn{1}{c}{0.973$\pm$0.01} & \multicolumn{1}{c}{0.869$\pm$0.07}& \multicolumn{1}{c}{0.840$\pm$0.07}& \multicolumn{1}{c|}{0.851$\pm$0.04} & \multicolumn{1}{c}{0.827$\pm$0.02}&\multicolumn{1}{c}{0.777$\pm$0.05} & \multicolumn{1}{c}{0.702$\pm$0.07}& \multicolumn{1}{c}{0.738$\pm$0.05}\\

     \multicolumn{1}{c}{Homogeneous motifs (All)} &\multicolumn{1}{c}{0.939$\pm$0.01} & \multicolumn{1}{c}{0.904$\pm$0.02}& \multicolumn{1}{c}{0.984$\pm$0.02}& \multicolumn{1}{c|}{0.942$\pm$0.01} & \multicolumn{1}{c}{0.797$\pm$0.01}&\multicolumn{1}{c}{0.749$\pm$0.01} & \multicolumn{1}{c}{0.911$\pm$0.01}& \multicolumn{1}{c}{0.817$\pm$0.01}\\

     \multicolumn{1}{c}{Heterogeneous motifs (All)} &\multicolumn{1}{c}{\textbf{0.987$\pm$0.01}} & \multicolumn{1}{c}{\textbf{0.976$\pm$0.02}}& \multicolumn{1}{c}{\textbf{1.000$\pm$0.00}}& \multicolumn{1}{c|}{\textbf{0.988$\pm$0.01}} & \multicolumn{1}{c}{0.833$\pm$0.02}&\multicolumn{1}{c}{0.802$\pm$0.02} & \multicolumn{1}{c}{0.893$\pm$0.02}& \multicolumn{1}{c}{0.842$\pm$0.01}\\

     \multicolumn{1}{c}{Heterogeneous motifs ($\text{AUC}^{\prime}_\text{upper}>0.7$)} &\multicolumn{1}{c}{\textbf{0.987$\pm$0.01}} & \multicolumn{1}{c}{\textbf{0.976$\pm$0.02}}& \multicolumn{1}{c}{\textbf{1.000$\pm$0.00}}& \multicolumn{1}{c|}{\textbf{0.988$\pm$0.01}} & \multicolumn{1}{c}{\textbf{0.833$\pm$0.01}}&\multicolumn{1}{c}{\textbf{0.803$\pm$0.02}} & \multicolumn{1}{c}{\textbf{0.894$\pm$0.01}}& \multicolumn{1}{c}{\textbf{0.841$\pm$0.01}}\\

     \hline
    \multicolumn{1}{c}{\multirow{2}*{Methods}}& \multicolumn{4}{c|}{TwiBot-20}& \multicolumn{4}{c}{TwiBot-22}\\
    \cline{2-9}
    \multicolumn{1}{c}{}&\multicolumn{1}{c}{Accuracy} & \multicolumn{1}{c}{Precision} & \multicolumn{1}{c}{Recall}& \multicolumn{1}{c|}{F1}&\multicolumn{1}{c}{Accuracy} & \multicolumn{1}{c}{Precision} & \multicolumn{1}{c}{Recall}& \multicolumn{1}{c}{F1}\\
    \hline
    \multicolumn{1}{c}{Ising \cite{des2022detecting}} &\multicolumn{1}{c}{0.573$\pm$0.01}& \multicolumn{1}{c}{0.587$\pm$0.03} & \multicolumn{1}{c}{0.857$\pm$0.02}& \multicolumn{1}{c|}{0.697$\pm$0.01}& \multicolumn{1}{c}{0.493$\pm$0.02}& \multicolumn{1}{c}{0.491$\pm$0.02} & \multicolumn{1}{c}{0.823$\pm$0.02}& \multicolumn{1}{c}{0.615$\pm$0.01} \\

    \multicolumn{1}{c}{SybilWalk \cite{jia2017random}} &\multicolumn{1}{c}{0.678$\pm$0.02}& \multicolumn{1}{c}{0.653$\pm$0.01} & \multicolumn{1}{c}{0.480$\pm$0.02}& \multicolumn{1}{c|}{0.553$\pm$0.03}& \multicolumn{1}{c}{0.537$\pm$0.01}& \multicolumn{1}{c}{0.537$\pm$0.02} & \multicolumn{1}{c}{1.000$\pm$0.00}& \multicolumn{1}{c}{0.699$\pm$0.02} \\

    \multicolumn{1}{c}{SybilSCAR \cite{wang2018structure}} &\multicolumn{1}{c}{0.473$\pm$0.02}& \multicolumn{1}{c}{0.400$\pm$0.03} & \multicolumn{1}{c}{0.466$\pm$0.01}& \multicolumn{1}{c|}{0.430$\pm$0.01}& \multicolumn{1}{c}{0.481$\pm$0.02}& \multicolumn{1}{c}{0.510$\pm$0.02} & \multicolumn{1}{c}{0.897$\pm$0.02}& \multicolumn{1}{c}{0.650$\pm$0.01} \\

    \multicolumn{1}{c}{Botometer \cite{yang2022botometer}} &\multicolumn{1}{c}{0.742$\pm$0.02}&\multicolumn{1}{c}{0.794$\pm$0.07} & \multicolumn{1}{c}{0.773$\pm$0.06}& \multicolumn{1}{c|}{0.783$\pm$0.04}& \multicolumn{1}{c}{0.750$\pm$0.01}&\multicolumn{1}{c}{0.820$\pm$0.06} & \multicolumn{1}{c}{0.837$\pm$0.06}& \multicolumn{1}{c}{0.825$\pm$0.04} \\

    \multicolumn{1}{c}{FP \cite{moghaddam2022friendship}} &\multicolumn{1}{c}{0.817$\pm$0.01}&\multicolumn{1}{c}{0.840$\pm$0.07} & \multicolumn{1}{c}{0.828$\pm$0.05}& \multicolumn{1}{c|}{0.831$\pm$0.03}& \multicolumn{1}{c}{0.811$\pm$0.01}&\multicolumn{1}{c}{0.830$\pm$0.06} & \multicolumn{1}{c}{0.841$\pm$0.08}& \multicolumn{1}{c}{0.832$\pm$0.05} \\

    \multicolumn{1}{c}{ARG \cite{abreu2020twitter}} &\multicolumn{1}{c}{0.803$\pm$0.02}&\multicolumn{1}{c}{0.834$\pm$0.08} & \multicolumn{1}{c}{0.830$\pm$0.08}& \multicolumn{1}{c|}{0.827$\pm$0.05}& \multicolumn{1}{c}{0.801$\pm$0.02}&\multicolumn{1}{c}{0.852$\pm$0.07} & \multicolumn{1}{c}{0.824$\pm$0.06}& \multicolumn{1}{c}{0.835$\pm$0.05} \\

    \multicolumn{1}{c}{DeeProBot \cite{hayawi2022deeprobot}} &\multicolumn{1}{c}{0.826$\pm$0.02}&\multicolumn{1}{c}{0.771$\pm$0.07} & \multicolumn{1}{c}{0.733$\pm$0.07}& \multicolumn{1}{c|}{0.751$\pm$0.07}& \multicolumn{1}{c}{0.833$\pm$0.02}&\multicolumn{1}{c}{0.783$\pm$0.06} & \multicolumn{1}{c}{0.841$\pm$0.08}& \multicolumn{1}{c}{0.807$\pm$0.04} \\

    \multicolumn{1}{c}{T5 \cite{raffel2020exploring}} &\multicolumn{1}{c}{0.819$\pm$0.02}&\multicolumn{1}{c}{0.729$\pm$0.07} & \multicolumn{1}{c}{0.867$\pm$0.06}& \multicolumn{1}{c|}{0.792$\pm$0.05}& \multicolumn{1}{c}{0.846$\pm$0.01}&\multicolumn{1}{c}{0.840$\pm$0.08} & \multicolumn{1}{c}{0.820$\pm$0.05}& \multicolumn{1}{c}{0.828$\pm$0.06} \\

    \multicolumn{1}{c}{BotRGCN \cite{feng2021botrgcn}} &\multicolumn{1}{c}{0.817$\pm$0.01}&\multicolumn{1}{c}{0.827$\pm$0.06} & \multicolumn{1}{c}{0.840$\pm$0.07}& \multicolumn{1}{c|}{0.833$\pm$0.05}& \multicolumn{1}{c}{0.851$\pm$0.01}&\multicolumn{1}{c}{0.715$\pm$0.08} & \multicolumn{1}{c}{0.808$\pm$0.07}& \multicolumn{1}{c}{0.759$\pm$0.05} \\

    \multicolumn{1}{c}{RGT \cite{feng2022heterogeneity}} &\multicolumn{1}{c}{0.844$\pm$0.01}&\multicolumn{1}{c}{0.720$\pm$0.09} & \multicolumn{1}{c}{0.892$\pm$0.07}& \multicolumn{1}{c|}{0.797$\pm$0.05}& \multicolumn{1}{c}{0.852$\pm$0.02}&\multicolumn{1}{c}{0.845$\pm$0.06} & \multicolumn{1}{c}{0.815$\pm$0.08}& \multicolumn{1}{c}{0.827$\pm$0.05} \\

    \multicolumn{1}{c}{Homogeneous motifs (All)} &\multicolumn{1}{c}{0.740$\pm$0.01}&\multicolumn{1}{c}{0.757$\pm$0.01} & \multicolumn{1}{c}{0.841$\pm$0.01}& \multicolumn{1}{c|}{0.791$\pm$0.01}& \multicolumn{1}{c}{0.844$\pm$0.00}&\multicolumn{1}{c}{0.844$\pm$0.00} & \multicolumn{1}{c}{0.839$\pm$0.00}& \multicolumn{1}{c}{0.839$\pm$0.00} \\

    \multicolumn{1}{c}{Heterogeneous motifs (All)} &\multicolumn{1}{c}{\textbf{0.825$\pm$0.01}}&\multicolumn{1}{c}{\textbf{0.810$\pm$0.01}} & \multicolumn{1}{c}{\textbf{0.923$\pm$0.01}}& \multicolumn{1}{c|}{\textbf{0.861$\pm$0.01}}& \multicolumn{1}{c}{\textbf{0.858$\pm$0.00}}&\multicolumn{1}{c}{\textbf{0.852$\pm$0.00}} & \multicolumn{1}{c}{\textbf{0.862$\pm$0.00}}& \multicolumn{1}{c}{\textbf{0.855$\pm$0.00}} \\

    \multicolumn{1}{c}{Heterogeneous motifs ($\text{AUC}^{\prime}_\text{upper}>0.7$)} &\multicolumn{1}{c}{0.757$\pm$0.01}&\multicolumn{1}{c}{0.747$\pm$0.01} & \multicolumn{1}{c}{0.891$\pm$0.01}& \multicolumn{1}{c|}{0.812$\pm$0.01}& \multicolumn{1}{c}{0.854$\pm$0.00}&\multicolumn{1}{c}{0.844$\pm$0.00} & \multicolumn{1}{c}{0.859$\pm$0.00}& \multicolumn{1}{c}{0.851$\pm$0.00} \\
    \hline
    \end{tabular}
    }
    \label{table:comsgb}
\end{table*}

\begin{table*}[htbp]
    \centering
\caption{Performance comparison of the AUC across different methods for social bot detection. Here, we report the average performance and standard deviation. For both homogeneous and heterogeneous motifs, results are obtained using a Gradient Boosting classifier. The optimal results are marked in bold.}
\begin{tabular}{cccccc}
\hline
Datasets & Cresci-15 & MGTAB & TwiBot-20 & TwiBot-22 \\
\hline
Ising \cite{des2022detecting} &0.553$\pm$0.02& 0.539$\pm$0.02 & 0.653$\pm$0.03 & 0.620$\pm$0.01 \\
SybilWalk \cite{jia2017random}  & 0.500$\pm$0.01 & 0.500$\pm$0.01 & 0.558$\pm$0.02 & 0.673$\pm$0.02 \\
SybilSCAR \cite{wang2018structure}  & 0.892$\pm$0.03 & 0.492$\pm$0.02 & 0.420$\pm$0.01 & 0.639$\pm$0.01\\
Botometer \cite{yang2022botometer} & 0.823$\pm$0.01 & 0.810$\pm$0.01 & 0.810$\pm$0.02 & 0.810$\pm$0.01 \\
FP \cite{moghaddam2022friendship} & 0.848$\pm$0.01 & 0.817$\pm$0.02 & 0.797$\pm$0.01 & 0.792$\pm$0.01 \\
ARG \cite{abreu2020twitter} & 0.838$\pm0.03$ & 0.807$\pm$0.01 & 0.809$\pm0.03$ & 0.813$\pm$0.02 \\
DeeProBot \cite{hayawi2022deeprobot} & 0.885$\pm0.01$ & 0.855$\pm$0.01 & 0.867$\pm$0.02 & 0.863$\pm$0.02 \\
T5 \cite{raffel2020exploring}& 0.855$\pm$0.01 & 0.835$\pm$0.01 & 0.841$\pm$0.02 & 0.857$\pm$0.01 \\
BotRGCN \cite{feng2021botrgcn}& 0.951$\pm$0.02 & 0.851$\pm$0.01 & 0.866$\pm$0.01 & 0.896$\pm$0.01 \\
RGT \cite{feng2022heterogeneity}& 0.915$\pm$0.01 & 0.842$\pm$0.02 & 0.848$\pm$0.02 & 0.862$\pm$0.02 \\
Homogeneous motifs (All)& 0.973$\pm$0.01 & 0.855$\pm$0.01 & 0.788$\pm$0.01 & 0.918$\pm$0.00 \\
Heterogeneous motifs (All)& \textbf{0.992$\pm$0.01} & \textbf{0.910$\pm$0.01} & \textbf{0.913$\pm$0.01} & \textbf{0.932$\pm$0.00} \\
Heterogeneous motifs ($\text{AUC}^{\prime}_\text{upper}>0.7$)& 0.987$\pm$0.01 & 0.909$\pm$0.01 & 0.853$\pm$0.01 & 0.928$\pm$0.00 \\
\hline
\end{tabular}
\label{table:aucgb}
\end{table*}

\subsubsection{Feature importance analysis of heterogeneous motifs}
\begin{figure}[htbp]
	\centering
	\resizebox{8.8cm}{!}{\includegraphics{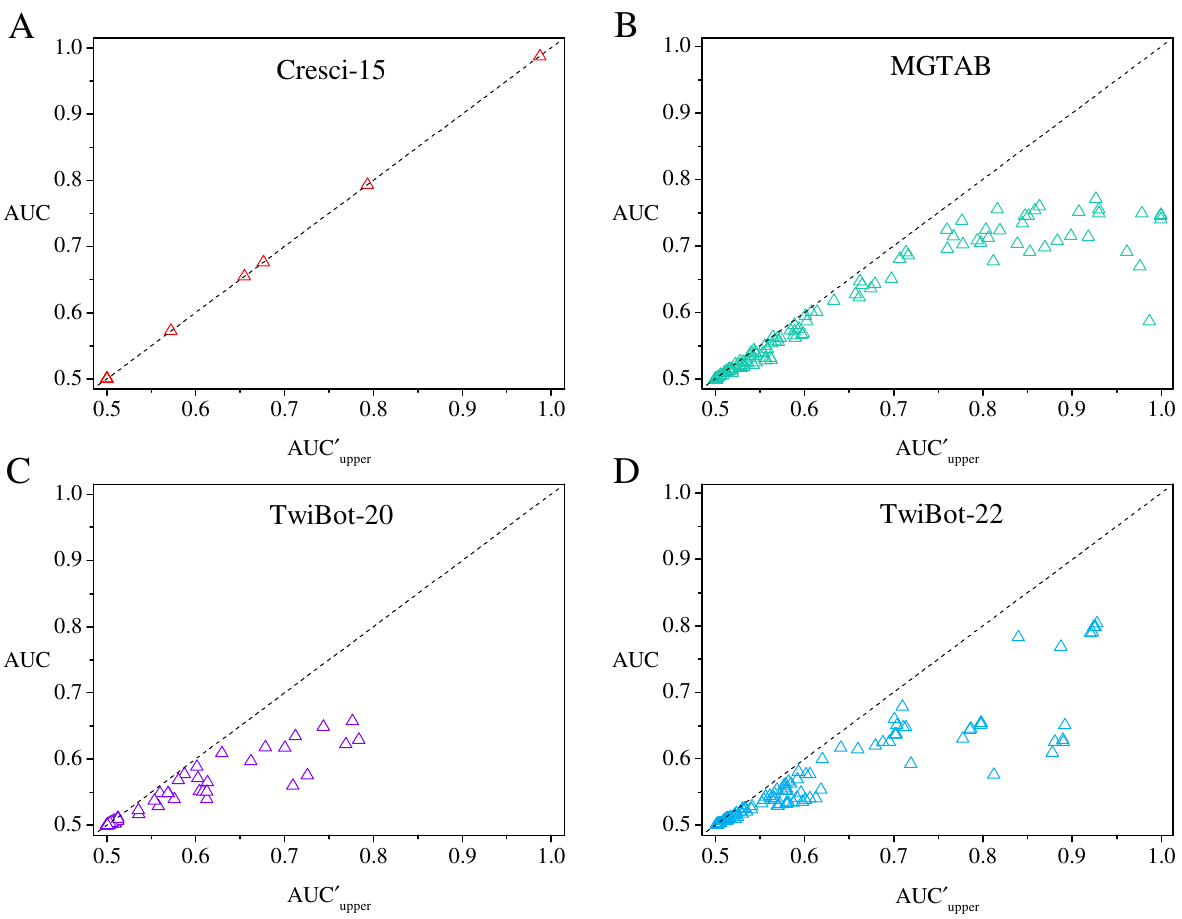}}
	\caption{The maximum capability of a heterogeneous motif. The horizontal axis is the maximum capability of a heterogeneous motif. The vertical axis is the real AUC performance of a heterogeneous motif by the XGBoost classifier. Note that each plot contains 114 data points. In the case of Cresci-15, most heterogeneous motifs have both a maximum capability and an actual AUC of 0.5, resulting in substantial overlap among the points.}
	\label{fig:mc}
\end{figure}

The results in Tables \ref{table:comsxgb} and \ref{table:aucxgb} indicate that our approach based on multiple heterogeneous motifs outperforms the 10 baselines in detection performance. However, it remains unclear which specific heterogeneous motifs play a key role in achieving this superior performance. To thoroughly assess the importance of heterogeneous motifs, we quantify the maximum capability of each feature to pre-evaluate which of the 114 heterogeneous motifs show strong promise. Specifically, we empirically determine the maximum capability of a heterogeneous motif on all four datasets (Fig. \ref{fig:mc}). Results indicate that some heterogeneous motifs demonstrate high capability, with $\text{AUC}^{\prime}_\text{upper}$ values exceeding 0.9 on MGTAB and TwiBot-22. However, selecting a single motif from these high-capability features ($\text{AUC}^{\prime}_\text{upper}>0.9$) for classifier training remains coarse-grained, limiting its effectiveness in distinguishing between humans and bots (with an actual $\text{AUC}<0.8$). As a result, the high capability of these motifs is not fully realized in classification.

Feature engineering and selection are critical to the success of binary classification tasks. In this study, we construct 114 heterogeneous motifs as engineered features, which have been shown to be effective for social bot detection (Tables~\ref{table:comsxgb}-\ref{table:aucgb}). However, not all features contribute equally to performance. Feature selection plays a key role in identifying the most informative subset of motifs to maximize detection accuracy. As shown in Fig.~\ref{fig:mc}, some heterogeneous motifs exhibit low capability. Therefore, focusing on motifs with higher capability and excluding less informative ones can lead to more accurate and efficient detection. 

To evaluate the impact of feature selection, we select a subset of heterogeneous motifs with $\text{AUC}^{\prime}_\text{upper} > 0.7$ for social bot detection. The results show that this subset achieves performance comparable to using all heterogeneous motifs—and even surpasses the full set on the Cresci-15 and TwiBot-22 datasets (Tables \ref{table:coms} and \ref{table:auc}). However, on the TwiBot-20 dataset, the selected subset performs slightly worse. This is primarily because the maximum capability of each heterogeneous motif on TwiBot-20 is relatively limited, with most values barely exceeding 0.8. It is important to note that while the actual AUC values of heterogeneous motifs vary across different classifiers, their maximum capability remains independent of the choice of machine learning algorithm. Overall, our theoretical framework not only quantifies the maximum capability of a heterogeneous motif but also serves as an effective feature selection mechanism for identifying a high-impact feature subset.

\section{Conclusion and Discussion}
\label{section6}
To summarize, we propose a novel approach to social bot detection based on a Na\"ive Bayes model, offering a theoretical framework to systematically quantify the contribution of different node pairs within heterogeneous motifs. By incorporating node-label information into homogeneous motifs, we construct heterogeneous motifs that effectively capture the diversity of neighborhood preferences, enabling a richer representation of local network structures. Extensive evaluations on four large-scale, publicly available benchmarks show that our method consistently outperforms state-of-the-art approaches across five standard evaluation metrics, demonstrating both robustness and generalizability. Additionally, we validate the effectiveness of the proposed features using Random Forest and Gradient Boosting, both of which yield performance comparable to that of our primary model. This confirms the adaptability of our framework across different classifiers. Finally, we introduce a theoretical method to quantify the maximum capability of a heterogeneous motif, showing that selecting only the most informative features can achieve near-optimal detection performance.

As large language models continue to advance, social bots are becoming increasingly sophisticated. They can easily fabricate user attributes and generate tweet content that is both semantically coherent and structurally complex, making content- and metadata-based detection increasingly challenging \cite{yang2023anatomy}. In response to this trend, our work focuses on a more robust and difficult-to-manipulate signal: the structural patterns of social interactions. To this end, we propose a theoretical framework based solely on network topology, using heterogeneous motifs to model neighborhood preference heterogeneity without relying on tweet content or user metadata. A promising direction for future work is to integrate the proposed heterogeneous motif-based features with other data modalities for social bot detection. These features offer complementary and discriminative structural signals that can enhance existing detection systems, particularly when combined with textual content and user profile attributes.

Detecting social bots on social platforms is both a significant and challenging task, particularly in the era of generative artificial intelligence \cite{yang2023anatomy}. Future research could benefit from incorporating additional structural information from networks, such as graphlets \cite{he2024uncovering}, hypergraphs \cite{purkait2016clustering}, and higher-order structures \cite{benson2016higher}, to address this issue more effectively. Another compelling avenue is the identification of group social bots by leveraging cluster behavior and collaborative behavior data \cite{pacheco2021uncovering}. In this study, we apply the Na\"ive Bayes model, which operates under the assumption that all features are conditionally independent given the target label. However, this conditional independence assumption may rarely hold in the real world \cite{jiang2009novel}. Consequently, exploring a Bayesian framework based on dependent assumptions could provide a valuable theoretical approach for advancing social bot detection. Taken together, our work presents an effective and theoretically grounded solution for social bot detection, offering a principled approach to feature selection while also opening avenues for future investigation into related research questions.

\section*{Acknowledgments}
This study's high-performance computations are supported by the Interdisciplinary Intelligence Supercomputer Center at Beijing Normal University, Zhuhai, People's Republic of China.

\section*{Code Availability}
The code used in this study is available at \href{https://github.com/YijunRan/Bot-detection}{https://github.com/YijunRan/Bot-detection}

\vfill

\end{document}